\documentclass[a4paper,12pt,useAMS,usenatbib]{mn2e}

\usepackage[table]{xcolor}
\usepackage{natbib}
\usepackage{float}
\usepackage{wrapfig}
\restylefloat{figure}
\usepackage{rotating}
\usepackage{amsmath}
\usepackage{graphicx}
\usepackage{multicol}
\usepackage{amssymb} 
\usepackage{multirow} 
\usepackage{epstopdf}
\usepackage{lscape} 
\usepackage{gensymb} 
\usepackage{siunitx} 
\usepackage{nicefrac} 
\usepackage{subfig} 
\usepackage{tabularx} 
\usepackage{booktabs}
\usepackage[flushleft]{threeparttable}
\usepackage{dblfloatfix} 
\usepackage{fixltx2e} 
\usepackage{placeins} 

\graphicspath{{../imgs/}}

\newcommand{\mytilde}{\raise.17ex\hbox{$\scriptstyle\mathtt{\sim}$}}

\title{The binary fraction of planetary nebula central stars: the promise of VPHAS+}
\author []
{\parbox{\textwidth}{Helen Barker$^{1\dagger}$\thanks{helen.barker@postgrad.manchester.ac.uk} , A. Zijlstra$^{1,2}$, Orsola De Marco$^{3,4}$, David J. Frew$^2$, J. E. Drew$^5$, R. L. M. Corradi$^7$, Jochen Eisl\"offel$^8$, Quentin A. Parker$^2$ }
	\vspace{0.6cm}\\
	\parbox{\textwidth}{
		$^1$Department of Astronomy and Astrophysics, The University of Manchester, Manchester, M13 9PL, UK\\
		$^2$Department of Physics \& Laboratory for Space Research, The University of Hong Kong, Hong Kong SAR, China \\
		$^3$Department of Physics \& Astronomy, Macquarie University, Sydney, NSW 2109, Australia\\
		$^4$Astronomy, Astrophysics and Astrophotonics Research Centre, Macquarie University, Sydney Australia\\
		$^5$ School of Physics, Astronomy \& Mathematics, University of Hertfordshire, College Lane, Hatfield, Hertfordshire, AL10 9AB, UK\\
		$^6$GRANTECAN, Cuesta de San Jos\'{e} s/n E-38712, Bre\~{n}a Baja, La Palma, Spain \\
		$^7$Instituto de Astrof\'{i}sica de Canarias, V\'{i}a L\'{a}ctea s/n, E38200, La Laguna, Tenerife, Spain \\
		$^8$  Th\"uringer Landessternwarte, Sternwarte 5, D-07778 Tautenburg, Germany \\
		$^\dagger$Visiting scholar, Macquarie University
}}

\begin{document}
\label{firstpage}

\maketitle

\begin{abstract}

The majority of planetary nebulae (PNe) are not spherical, and current single-star models cannot adequately explain all the morphologies we observe. This has led to the Binary Hypothesis, which states that PNe are \emph{preferentially} formed by binary systems. This hypothesis can be corroborated or disproved by comparing the estimated binary fraction of all PNe central stars (CS) to that of the supposed progenitor population. One way to quantify the rate of CS binarity is to detect near infra-red (IR) excess indicative of a low-mass main sequence companion. In this paper, a sample of known PNe within data release 2 of the ongoing VPHAS+ are investigated. We give details of the method used to calibrate VPHAS+ photometry, and present the expected colours of CS and main sequence stars within the survey. Objects were scrutinized to remove PN mimics from our sample and identify true CS. Within our final sample of 7 CS, 6 had previously either not been identified or confirmed. We detected an $i$ band excess indicative of a low-mass companion star in 3 CS, including one known binary, leading us to to conclude that VPHAS+ provides the precise photometry required for the IR excess method presented here, and will likely improve as the survey completes and the calibration process finalised. Given the promising results from this trial sample, the entire VPHAS+ catalogue should be used to study PNe and extend the IR excess-tested CS sample.

\newpage
\end{abstract}


\newpage

\section{Introduction}
\label{sec:introduction}

The precise origin of the shaping of planetary nebulae (PNe) remains uncertain. Observations find that approximately 80\% of PNe are non-spherical, with the majority of these exhibiting slight ellipticy or localised asymmetries. These mildly aspherical morphologies can be explained by interaction of the PN wind with the interstellar medium (ISM), or via the Generalised Interacting Stellar Wind (GISW) model (eg. \citealt{kwok1978}, \citealt{balick&frank2002}), which invokes a dense equatorial torus to funnel the stellar winds. However, it is unclear how such an equatorial torus could form, and there is no single star model that generates the required rotation rates and magnetic fields to form highly elliptical or bipolar PN morphologies \citep{garciasegura2014}.

A second model of PN formation requires an orbiting binary companion to shape the central star's (CS) ejecta via the gravitational interaction or via the generation of rotation and magnetic fields, allowing the formation of a wider range of aspherical morphologies. Given the prevalence of aspherical PNe, this second model has been expanded into the Binary Hypothesis \citep{demarco2009}, which states that observable PNe are \emph{preferentially} formed by binary systems.

It is generally accepted that CS binarity is an import factor in the shaping of PNe, (e.g., \citealt{miszalski2009b}, \citealt{hillwig2016}), but there is evidence that binary evolution also impacts other physical properties of the PN. For example, \cite{frew&parker2007} suggested that post-common envelope (CE) PNe, where a companion has spiralled in due to interaction with the asymptotic giant branch (AGB) star's ejecta to form a very close binary, have, on average, lower ionised masses. \cite{frew2016} found that known PNe with close binary CS have a restricted range of H$\alpha$ surface brightnesses, and suggested that this is a physical rather than a selection effect as previously interpreted \citep{bond&livio1990}. \cite{corradi2015} observed several post-CE binaries, and concluded that the large difference between the abundances calculated using the optical recombination lines and the collisionally excited lines, known as the abundance discrepancy factor, is related to the binary stellar evolution. This has since been supported by other observations (\citealt{wesson2016}, \citealt{jones2016}). Therefore, if the Binary Hypothesis is valid, there will be implications for any calculation which has assumed single-star PN formation, such as studies of stellar evolution, molecule formation, and ISM enrichment.

One way to test the Binary Hypothesis is to compare the CS binary fraction to the binary fraction of the progenitor main sequence population, measured to be $(50\pm4) \%$ by \cite{raghavan2010}. Alas, detecting CS binarity is not trivial. While detecting binaries with periods shorter than a few days is relatively easy (e.g., \citealt{hillwig2017}, \citealt{jones2015}), high precision observations and careful reduction techniques are required for wider binaries (e.g., \cite{hrivnak2017}). This means that sample sizes are small, making statistical analysis of the total PN population difficult.

Here, we use the infra-red (IR) excess method, described fully in Section \ref{sec:ir_excess_method}, to search for binaries in PNe. This builds upon the work of \cite{demarco2013} and \cite{douchin2015}, who attempted to constrain the CS binary fraction by searching for \emph{I} and \emph{J} band flux excess in samples of CS within the volume-limited sample of \cite{frew2008}. Small number statistics made it impossible to accurately account for observational biases and constrain the total PN binary fraction, so \cite{douchin2015} also used archival Sloan Digital Sky Survey (SDSS) data to search for $z$ band excess. It was concluded that although SDSS photometry is intrinsically sufficiently precise to detect IR flux excess, other limitations such as the inability to subtract bright nebulae resulted in too high an uncertainty, on average, to extend the PN sample.

Here, we examine a trial sample of PNe within the footprint of the VLT Survey Telescope (VST) H$\alpha$ Survey of the Southern Galactic Plane and Bulge (VPHAS+) \citep{drew2014}, to determine whether the new high-resolution images will allow us to identify new CS candidates, and if the photometry is sufficiently precise enough to detect CS IR excess. If successful, the techniques developed in this work can be applied to all the known PNe within the survey footprint, giving a much larger sample from which the CS binary fraction can be determined.

The specification of VPHAS+ and calibration performed are summarised in Section \ref{sec:vphas}. The sample of known PNe within VPHAS+ with identified CS or strong candidates are presented in Section \ref{sec:pn_sample}. The results of the investigation into CS binarity are presented in Sections \ref{sec:analysis} and \ref{sec:results_excesses}, with notes on individual objects presented in Section \ref{sec:individual_pn}. We conclude in Section \ref{sec:conclusion}.

\section{IR excess method}
\label{sec:ir_excess_method}

Although low-mass main sequence stars, the most common type of companion in binary CS, are relatively dim compared to the CS itself, they can be detected in a spatially unresolved binary system as the CS radiates primarily at blue and UV wavelengths, and the main sequence star radiates mostly at red and IR wavelengths. To detect binary CS in practice, observations in at least two blue bands and one IR band are required. In this work, the two blue bands used are the Sloan $u$ and $g$ bands, and the IR band is the Sloan $i$ or 2MASS $J$ band. The $J$ band is more sensitive to cool companions, but is not included in VPHAS+ data, so literature $J$ band measurements are used where sufficiently precise values were available. The IR-excess detection method is summarised as follows:
\begin{enumerate}
	\item We adopt a model stellar atmosphere of an appropriate effective temperature to represent a given single CS. The temperature is either obtained from the literature, a value adopted following a Zanstra analysis (see \cite{pottasch1984} for a description of the method) and other considerations, or is assumed to be 100kK.
	\item We compare the modelled to the observed $u-g$ colour to determine the reddening, $E(B-V)$, of the CS.
	\item We de-redden all observed magnitudes using this $E(B-V)$ value.
	\item We compare the modelled $g-i$ (and $g-J$ where available) colour to the observed, de-reddened colour. Any excess in the observed de-reddened colour is interpreted as a red or IR excess, possibly indicative of a low mass main sequence companion.
\end{enumerate}

It should be noted that this method can only select binary \emph{candidates}. Follow-up observations are required as IR excess can come from other sources, such as a non-standard stellar atmosphere, or simple line-of-sight superposition of stars.

\begin{figure}
	\begin{center}
		\includegraphics[width=\linewidth]{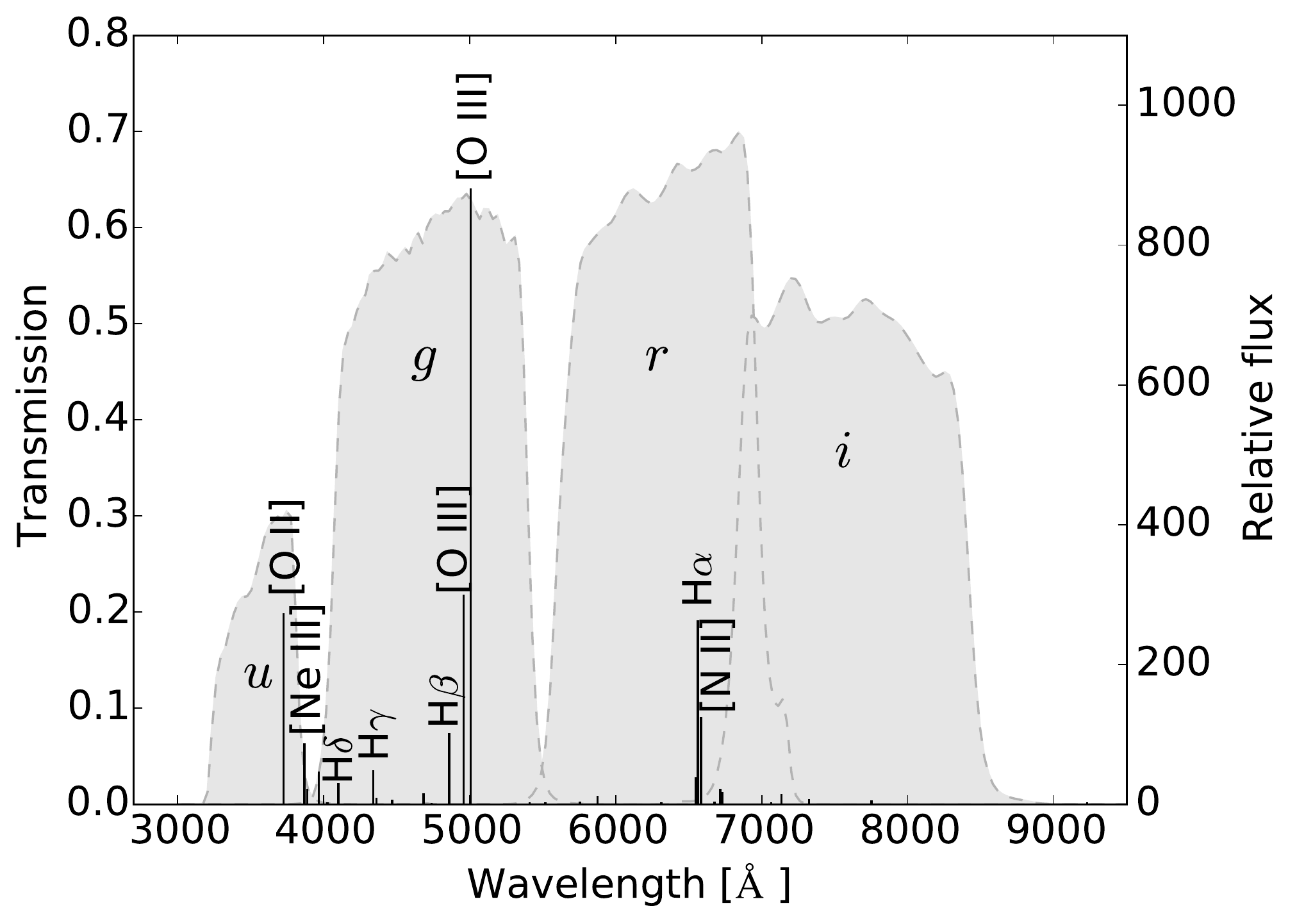}
		\caption{The Sloan bands overlaid the spectrum of a typical unreddened PN. }
		\label{fig:pn_lines_in_bands}
	\end{center}
\end{figure}

Another source of uncertainty can also come from nebula contamination, the effects of which are illustrated in Figure \ref{fig:pn_lines_in_bands}. The transmission windows of the Sloan $u$, $g$, $r$ and $i$ bands are shown, overlaid with a typical, un-reddened PN spectrum, that of NGC~3587 \citep{kwitter2001}. We 

\begin{figure*}
	\begin{center}
		\includegraphics[width=\linewidth, height=5.6cm, keepaspectratio ]{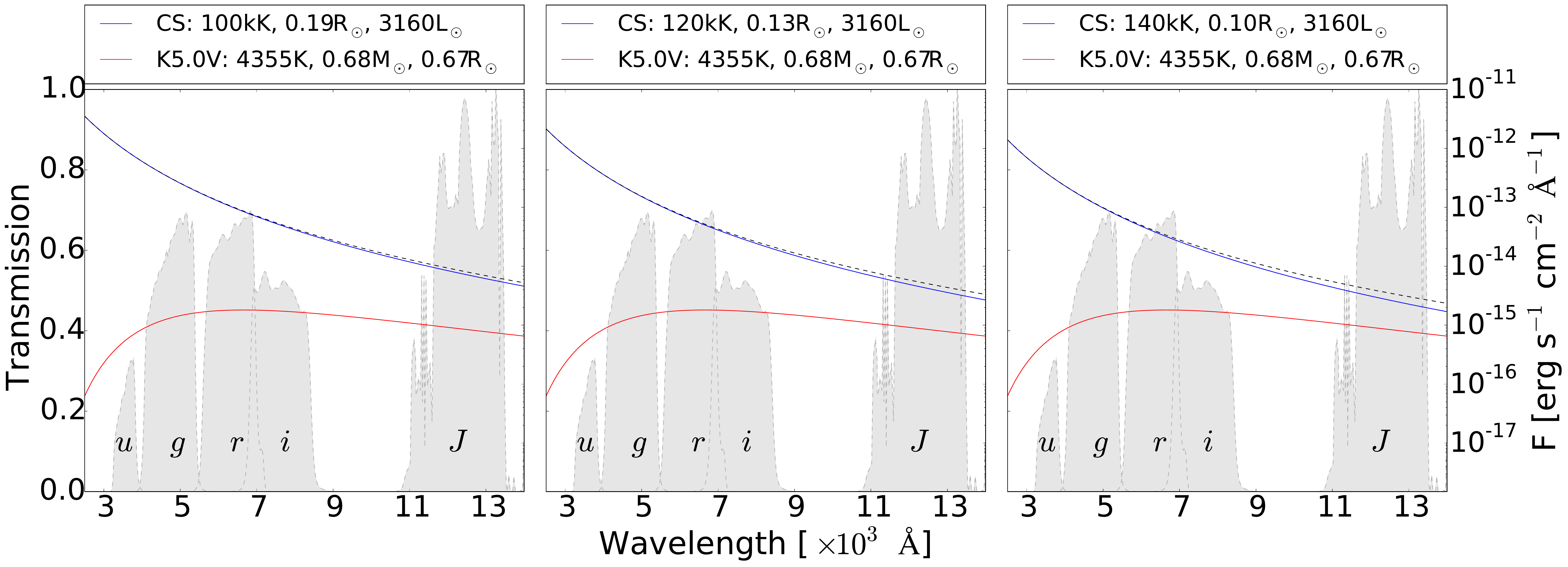}
		\caption{Blackbody curves of CS on the horizontal track (blue solid line) with a luminosity of 3160L$_{\odot}$ and effective temperatures of 100kK (left panel), 120kK (middle panel), and 140kK (right panel), and a relatively bright K5V main sequence star (red line) modelled using observations from \protect\cite{boyajian2012}. All stars are de-reddened and placed at 1 kpc. The dashed line showing the combined flux of the system is discernible from the blue line of the CS only at red wavelengths.
		}
		\label{fig:bb_horizontaltrack_excess}
	\end{center}
\end{figure*}

\noindent
can see an [OII] line in the $u$ band, [OIII] and H$\beta$ lines in the $g$ band, and H$\alpha$ and [NII] lines in the $r$ band. In most cases, we would expect [OIII]$\gg$[OII], H$\alpha$ and [NII].

The $u-g$, rather than the $u-r$ colour, is used to calculate $E(B-V)$ as it should be least affected by the presence of a possible low-mass companion. However, a strong [OIII] line that is not properly subtracted would make the $u-g$ colour redder than it should be. This extra `redness' would be attributed to interstellar reddening, erroneously increasing the calculated $E(B-V)$ value and causing any companion's contribution to the overall flux to be reduced in the de-reddening process. Nebula contamination can therefore lead to missed detections of binary companions (false negatives). To avoid this, large, dispersed, and faint PNe observed under good seeing conditions are required for this method, since accurate nebula light subtraction is almost impossible to achieve for bright and/or compact PNe.

The greatest limitation of the IR excess detection method is the requirement for extremely precise CS photometry (and PN subtraction for bright PNe), as the flux emitted by a putative low-mass companion is very small both intrinsically and in comparison to the highly luminous CS. On the ``horizontal track'', the transition at constant luminosity between the AGB and white dwarf (WD) cooling track, CS are so luminous that they outshine all but the brightest main sequence companions. Figure \ref{fig:bb_horizontaltrack_excess} shows spectral energy distributions representative of CS on the horizontal track with a fixed luminosity of 3160L$_{\odot}$, a mass of 0.6M$_{\odot}$, at a distance of 1kpc (no reddening), and temperatures of 100kK, 120kK and 140kK. Also shown is the spectrum of a K5V star, with mass, radius and temperature parameters from \cite{boyajian2012}. All objects are assumed to radiate as blackbodies. The dashed lines show the combined flux of the binary system and the filter curves of the Sloan $u$, $g$, $r $ and $i$ bands, and 2MASS $J$ band are overlaid. The relatively bright K5V star makes only a small contribution to the total flux, and is only discernible from the CS at the redder wavelengths.

\begin{figure*}
	\begin{center}$
		\begin{array}{c}
		\includegraphics[width=\linewidth, height=5.6cm, keepaspectratio]{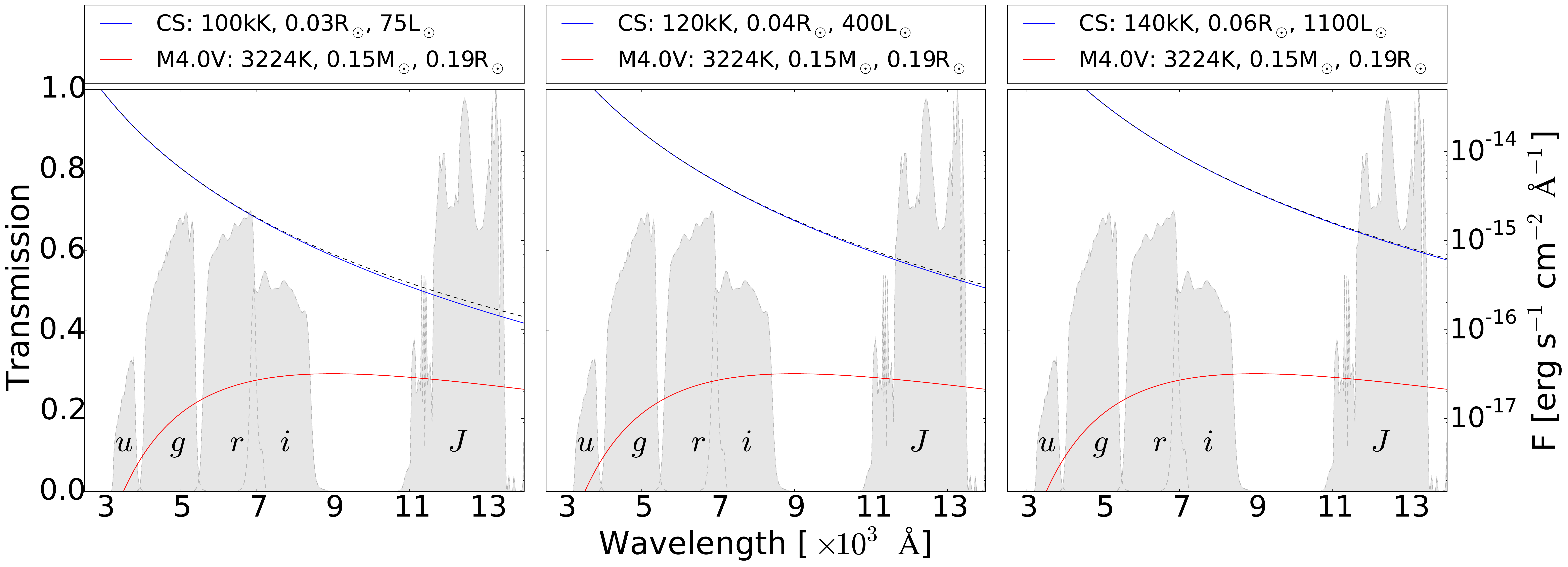}\\
		\includegraphics[width=\linewidth, height=5.6cm, keepaspectratio]{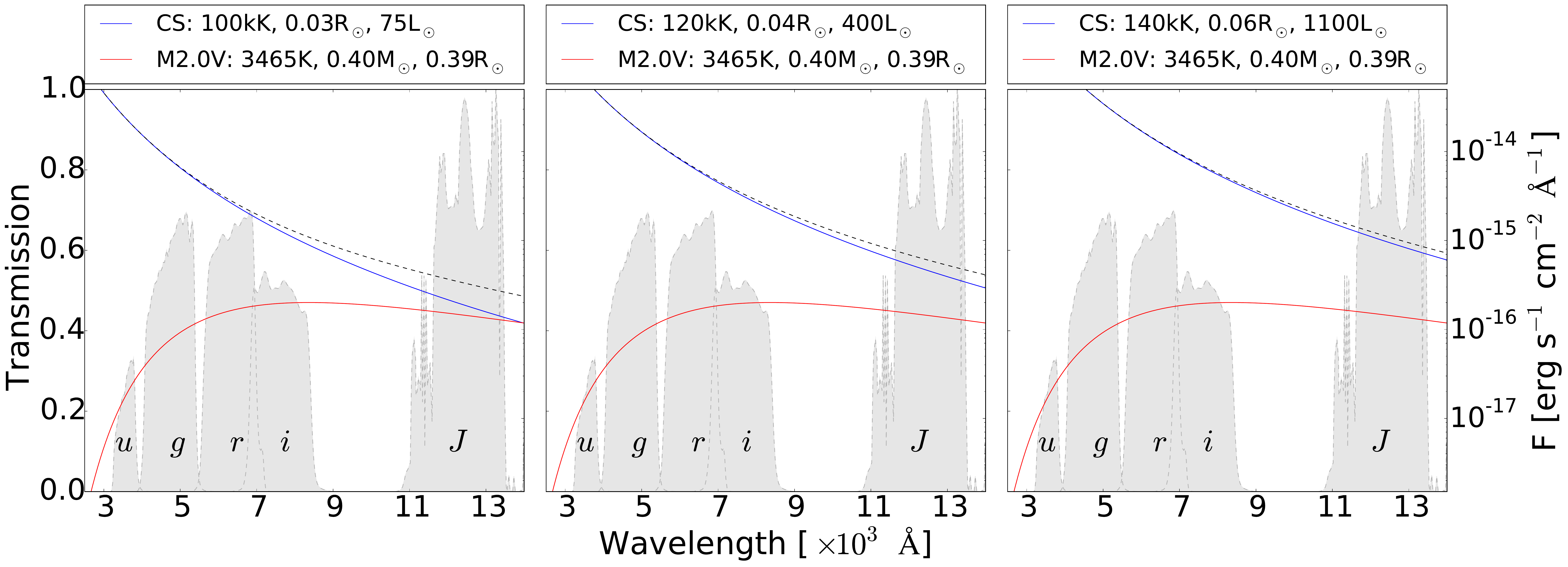}\\
		\includegraphics[width=\linewidth, height=5.6cm, keepaspectratio]{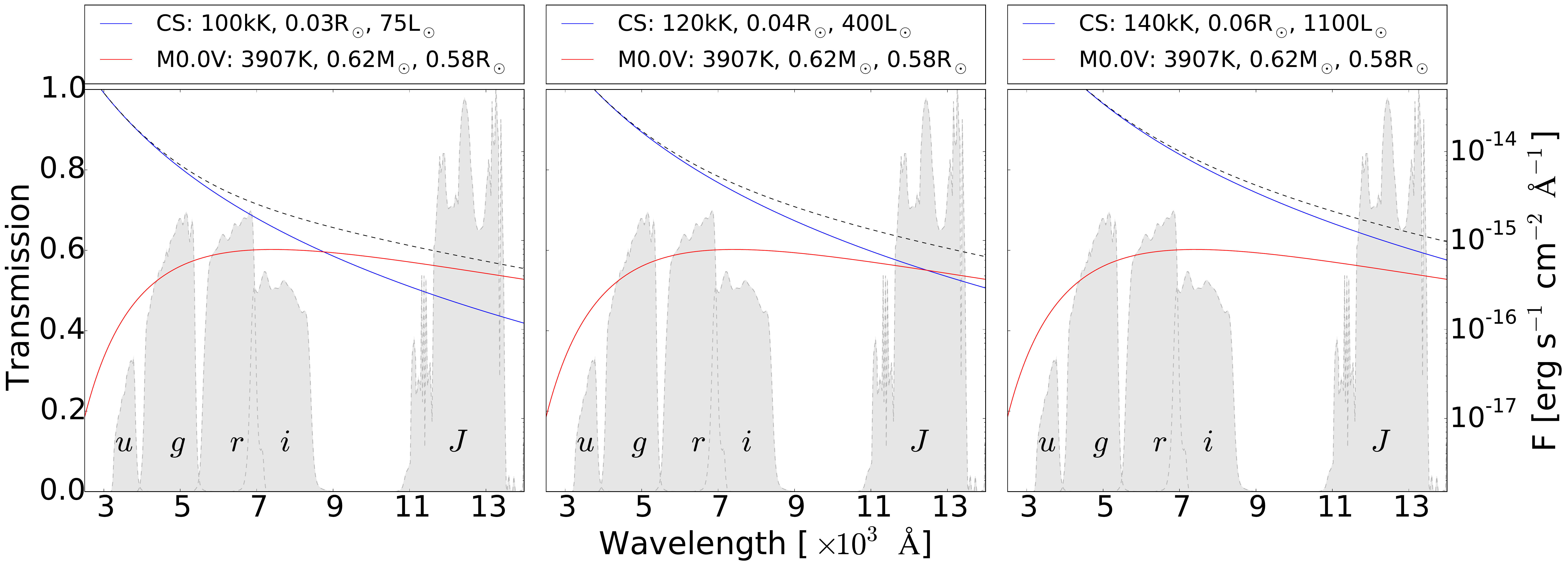}\\
		\includegraphics[width=\linewidth, height=5.6cm, keepaspectratio]{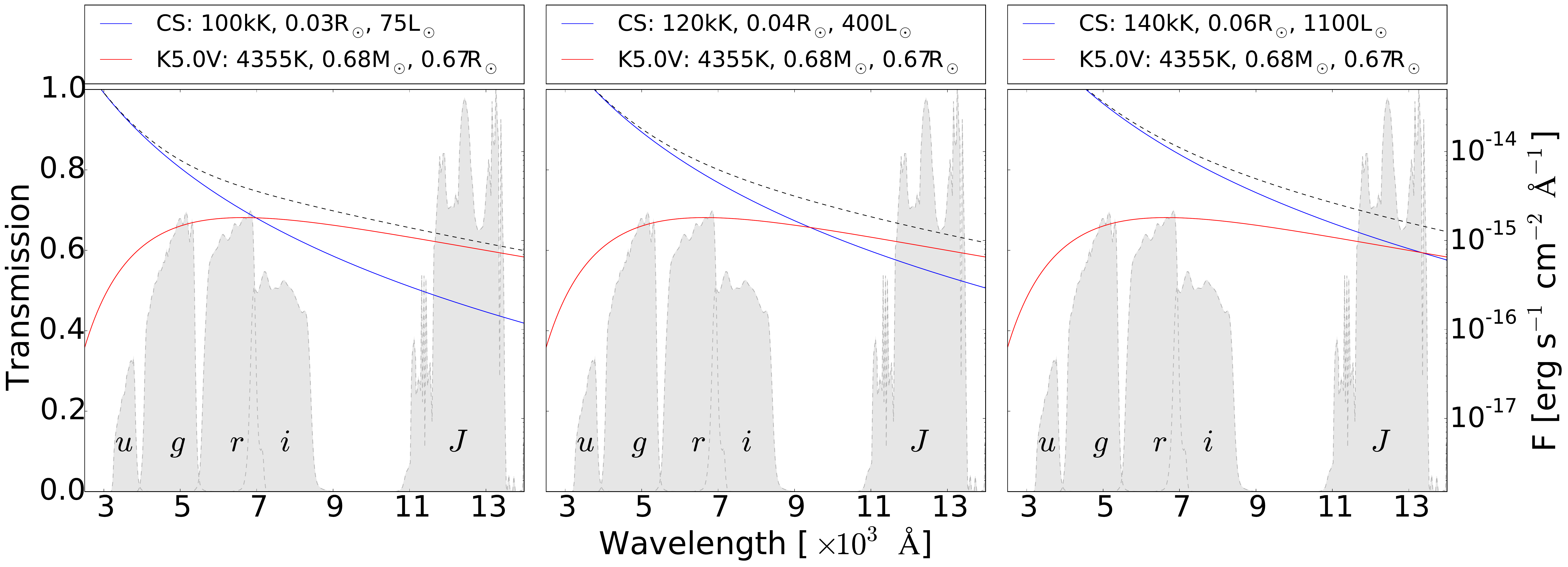}\\
		
		\end{array}$
		\caption{Blackbody curves of CS on the cooling track (blue solid line), modelled using temperature, radius and luminosity values from \protect\cite{vassiladiswood1994}, and main sequence companions (red line) modelled using physical parameters from \protect\cite{boyajian2012}. The dashed line shows the combined flux of the pairs.}
		\label{fig:bb_excess}
	\end{center}
\end{figure*}

Figure \ref{fig:bb_excess} shows the spectra of CS on the cooling track with temperatures of 100kK, 120kK and 140kK using mass and radius parameters from \cite{vassiladiswood1994} and the spectra of a range of possible low-mass main sequence companions \citep{boyajian2012}. As CS luminosity decreases with temperature on the cooling track, lower mass main sequence companion stars become distinguishable from the CS. These plots also allow us to identify another limitation of the IR excess detection method; a K5 star is blue enough that its flux contaminates the $g$ band.  Similar to the effect of nebula contamination, this $g$ band contamination will lead to a larger-than-real $E(B-V)$, leading in turn to dereddened colours that are too blue, resulting in a reduction of the red/IR flux excess. In this way, the K5 star may be entirely removed during the de-reddening process, making its detection impossible.  It should also be noted that for spectral types earlier than approximately G0V, the analysis presented here simply does not work, as the G star's colours, reddened by interstellar dust, simply dominate and our assumption of the intrinsic CS colours fails. However, the complete breakdown of our technique serves in itself to indicate the presence of a bright companion, which can then be detected and confirmed with alternative techniques.

\begin{figure}
	\centering
	\includegraphics[width=\linewidth]{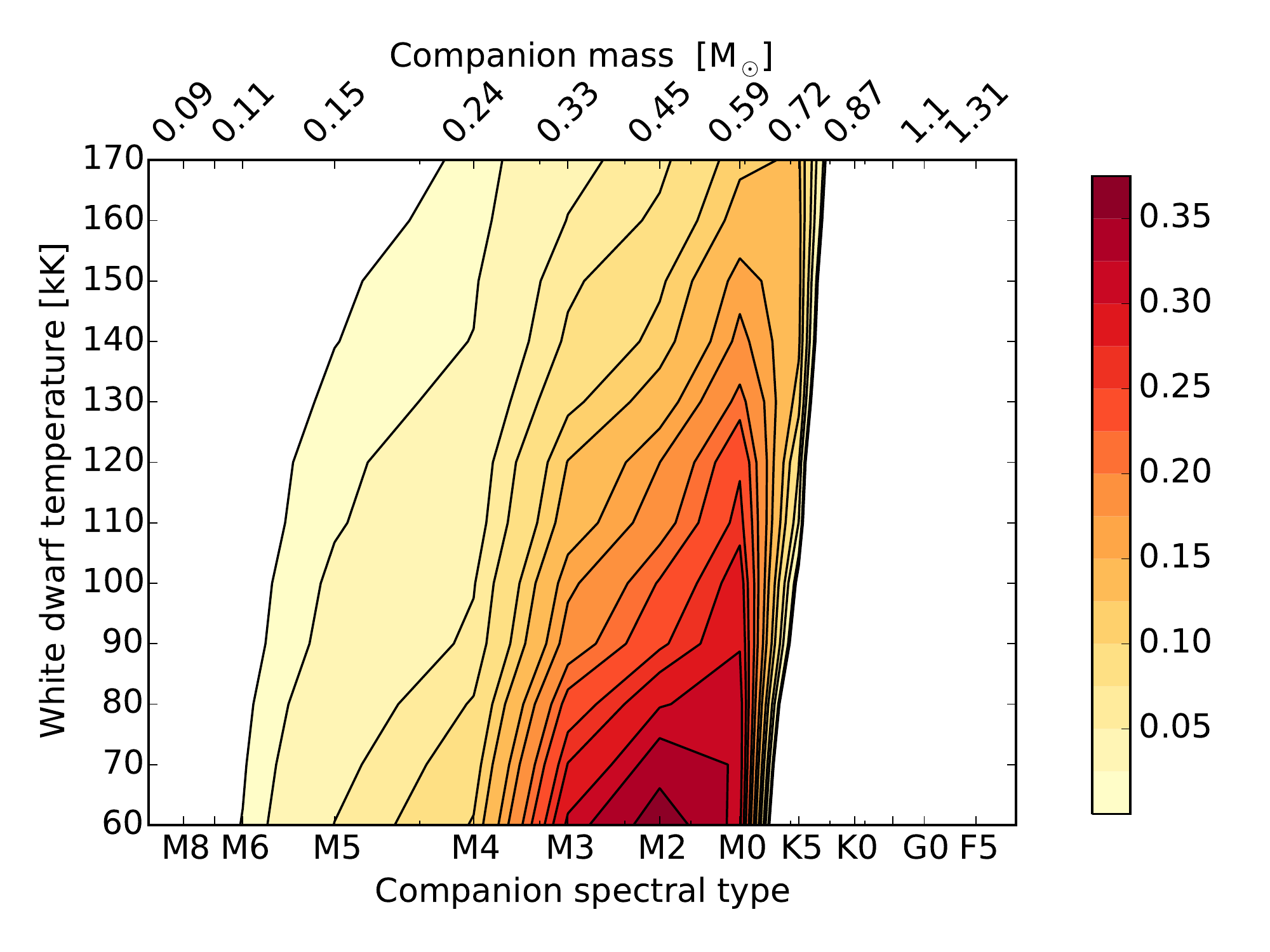}
	\caption{Contour plot showing the expected observed $i$ band excess for CS binary systems (in magnitudes), assuming the CS is on the cooling track, and accounting for over de-reddening caused by contamination of the $g$ band by the companion main sequence star.}
	\label{fig:contour_plot}
\end{figure}

Figure \ref{fig:contour_plot} allows us to visualise the combined effects of low companion luminosity and $g$ band contamination. A grid of binary system colours were calculated for a CS on the white dwarf cooling track and a range of main sequence companion spectral types. The $u-g$ colours of each CS - companion combination were compared to those of the CS only, and the difference attributed to reddening. The reddening value derived from this comparison was used to de-redden all the bands of the synthetic binary. The $g-i$ colour of the binary system was then compared to that of the CS alone, and any difference attributed to an excess of flux in the $i$ band. 

The calculated $i$ band excess is shown as a contour plot. The bunched contours around an M0V companion spectral type show the quickly decreasing $i$ band excess caused by the $g$ band contamination. The precision photometry required for this technique can be appreciated; even under ideal conditions with a low temperature faint CS and a late M companion star, we only expect an $i$ band excess of a few hundredths of a magnitude.

\section{VPHAS+}
\label{sec:vphas}

\subsection{Overview}
\label{subsec:vphas_overview}

VPHAS+ is an ESO public survey surveying the Galactic Plane south of the celestial equator at latitudes $\lvert \emph{b} \rvert<5 \degree$, extending to $\lvert \emph{b} \rvert<10 \degree$ across the Galactic Bulge, providing photometry down to $\sim$20th magnitude in the Sloan $u$, $g$, $r$, and $i$ bands, and in an additional H$\alpha$ narrowband. The median seeing in $g$ is 0.8 arcsec and $\sim$1.0 arcsec in $u$, sampled at 0.21 arcsec per pixel by OmegaCAM. The field of view is one square degree and is imaged onto a mosaic of 32 CCDs. 

Observations are split into red  and blue observing blocks to ensure the more stringent $u$ band observing conditions are met whilst allowing the survey to progress in a timely manner. The red block consists of $r$, $i$ and H$\alpha$ exposures. The blue block contains a second $r$ band exposure, labelled $r$2, and exposures in the $u$ and $g$ bands. As well as being used to check the stability of VPHAS+ photometry, the repeated $r$ band measurements provide an independent test for CS variability. In this work, we have used data from VPHAS+ data release 2 (DR2), which covers 24\% of the total survey footprint.

To minimise the survey footprint gaps due to the spaces between the CCDs and reduce the impact of poor photometry at CCD edges, every VPHAS+ pointing comprises of several offsets. Two of these offsets comprise of an exposure in every filter; these are referred to as block A and block B. The third offset, referred to in this work as Block C, comprises of a $g$ and H$\alpha$ exposure.


\subsection{Calibration}
\label{subsec:vphas_calibration}

The data used throughout this work was acquired from the Cambridge Astronomical Survey Unit (CASU)\footnote{http://casu.ast.cam.ac.uk/}, where a pipeline checks and processes raw VPHAS+ data; see \cite{drew2014} for details. The pipeline produces a single-band catalogue in each filter for every CCD in a VPHAS+ pointing. The catalogues contain flux counts for objects using a range of apertures. A summary of the radii of these apertures is provided in Table \ref{tab:aperture_radii}, along with the fraction of the point spread function (PSF) inside this aperture for a seeing of 0.8 arcseconds. 

\begin{table}
	\begin{center}
		\caption{Aperture numbers and radii provided in the VPHAS+ single band catalogues from CASU, and the fraction of the PSF inside the aperture if the seeing is 0.8 arcseconds.}
		\scalebox{0.8}
		{\begin{tabular}{c c c}
				\hline
				Aperture & Radius & Fraction \\
				number & (arcseconds) & of flux \\
				\hline
				1 & 0.5 & 0.41 \\
				2 & $\nicefrac{1}{\sqrt{2}}$ & 0.58 \\
				3 & 1.0 & 0.74 \\
				4 & $\sqrt{2}$ & 0.86 \\
				5 & 2.0 & 0.93 \\
				6 & 2$\sqrt{2}$ & 0.97\\
				7 & 4.0 & 1.0 \\
				8 & 5.0 & 1.0\\
				\hline
		\end{tabular}}
		\label{tab:aperture_radii}
	\end{center}
\end{table}

\subsubsection{Calibrating the $g$, $r$ and $i$ bands}	
\label{subsubsec:calibrating_gri}	

As a by-product of pipeline illumination correction, VPHAS+ magnitudes are referred to the photometric scale of the American Association of Variable Star Observers (AAVSO) Photometric All-Sky Survey (APASS), which provides magnitudes in the AB system. This procedure directly calibrates the Sloan $g$, $r$ and $i$ bands common to both APASS and VPHAS+, using two-dimensional functional fits spanning the full 32-CCD detector footprint.

Although the VPHAS+ pipeline provides aperture corrections to move VPHAS+ magnitudes onto a standard scale, they were not used in this work as a global VPHAS+ calibration has not yet been finalised, and regardless, the high precision photometry required here merits its own calibration process. The remainder of this section describes how we calibrated photometry from the VPHAS+ single-band catalogues.

To match in-common stars within a VPHAS+ CCD to APASS, VPHAS+ detected stellar objects falling within a 1.5 arcsecond radius of each APASS star with more than one observation were examined. The brightest VPHAS+ star in each instance was selected as the true match, reducing the number of mismatches due to inaccurate astronometry in APASS. Matches where the star was flagged as saturated, brighter than 13th magnitude or fainter than 16th magnitude in the VPHAS+ catalogue were then removed, leaving only those with the best photometry. This magnitude range is limited so as to avoid saturation in VPHAS+, and poor faint star photometry in APASS.

For every object on this list of APASS-VPHAS+ matches, VPHAS+ AB magnitudes were calculated using each of the pipeline-generated aperture fluxes.  A new customised aperture correction was then calculated to account both for any shift in overall zero point, and for flux not enclosed by the aperture. The aperture correction used is equal to the mean magnitude difference between APASS and VPHAS+ measurements for stars between 13th and 16th magnitudes, with the error equal to the standard deviation.

Using this calibration method, VPHAS+ magnitudes measured using the 1.0 arcsecond radius aperture typically have systematic errors less than 0.01 magnitudes, and formal errors on single measurements of a $\sim$19$^{th}$ magnitude object are less than 0.016, 0.025, and 0.044 in the $g$, $r$ and $i$ bands, respectively. The uncertainty on $i$ band measurements is greater than for the $g$ and $r$ bands as there is more noise associated with the photon count. These formal errors, of course, decreases for objects with multiple measurements, and there are usually two or more detections available.

\subsubsection{Calibrating the $u$ band}
\label{subsubsec:calibrating_u}

The VPHAS+ $u$ band was calibrated as described by \cite{drew2014}, by plotting ($u-g$) vs ($g-r$) colour-colour diagrams and overlaying synthetic stellar reddening tracks. These tracks are provided in the Vega magnitude system, so VPHAS+ magnitudes were first converted using: 
\begin{equation}
m_{\text{AB}} = m_{\text{Vega}} + \text{offset},
\end{equation}
where the offset is the ratio of the flux zero-points, $F_{zp}$, of the filter, in Janskys, for each magnitude system :
\begin{equation}
\text{offset} = 2.5 \log_{10}\left(\frac{F_{zp, \text{AB}}}{F_{zp, \text{Vega}}}\right)
\end{equation}

The flux zero-point of the AB magnitude system is defined to be 3631.0 Jy. The zero-points of the VPHAS+ filters in the Vega system were obtained from the Spanish Virtual Observatory (SVO) website\footnote{http://svo2.cab.inta-csic.es/svo/theory/fps3/}. The calculated Vega to AB corrections for VPHAS+ are shown in  Table \ref{tab:apass_corrections}. It should be noted that the AB-Vega offset in the $u$ band is almost invariably too large by up to 0.2 magnitudes, but this is corrected for by the calibration process. Vega magnitudes are used throughout the rest of this work.

\begin{table}
	\begin{center}
		\caption{The offset values used to convert between the Vega and AB magnitude system in VPHAS+. The magnitude offset in the $u$ band is likely too large by up to 0.2 magnitudes, but is accounted for during the calibration process.}
		\scalebox{0.8}
		{\begin{tabular}{lccc}
				\hline
				VPHAS+ & Magnitude & Vega & AB \\ 
				filter & offset & zero-point (Jy) &  zero-point (Jy) \\
				\hline      
				$u$ & 0.924* & 1550.81 & 3631.00 \\
				$g$ & -0.094 & 3960.53 & 3631.00 \\
				$r$ & 0.174 & 3094.68 & 3631.00 \\
				$i$ & 0.378 & 2563.84 & 3631.00 \\
				\hline
				\multicolumn{4}{l}{* Likely too large by up to 0.2 magnitudes}	\\	
		\end{tabular}}
		\label{tab:apass_corrections}		
	\end{center}
\end{table}

The calibrated single-band CCD catalogues were then combined to create a band-merged catalogue for each block in a VPHAS+ pointing, and ($u-g$) vs ($g-r$) colour-colour diagrams plotted using the calibrated $g$ and $r2$ Vega magnitudes. The $u$ band Vega magnitudes were adjusted until the maximum number of stars lay between the un-reddened main sequence and G0V reddening lines.

This method of calibrating the $u$ band results in larger systematic magnitude errors; typically 0.025 magnitudes using the 1.0 arcsecond aperture, with a typical total error on a single measurement of a $\sim$19$^{th}$ magnitude object of 0.048 magnitudes.

\subsection{Predicted colours of central stars in VPHAS+}
\label{subsec:predicted_colours}
The expected colours and magnitudes on the Vega system of both CS and main sequence stars when observed in VPHAS+ were calculated. First, the VPHAS+ bandpasses and an additional $J$ band were constructed, because although the VPHAS+ bandpasses are based on those of SDSS, they are not identical, and OmegaCAM is more sensitive to blue wavelengths than the detector used by SDSS. The constructed bandpasses included CCD response and atmospheric extinction at Paranal \citep{patat2011}. Spectra from the literature were then convolved with these bands, as described below.

\subsubsection{Colours of single central stars}
\label{subsec:cspn_predicted_colours}

To calculate the expected colours and magnitudes of CS in VPHAS+, synthetic spectral energy distributions (SEDs) from the TMAP database were used: theoretical NLTE model stellar atmospheres calculated using the code TMAW \citep{rauch2003, werner1999, werner2003}. The chosen models have a range of effective temperatures $T_{\text{eff}}$ from 20kK to 170kK, with log(g) values in the range 4.0 to 8.0. Colours calculated for each of these models are listed in Table \ref{tab:synthetic_CS_colours}. Here, it can be seen that the colours are almost independent of the choice of surface gravity, and in the hotter models, almost independent of temperature.

The effect of reddening on CS colours was then investigated. The spectrum of the $T_{\text{eff}}$=100kK, log(g)=7.0 TMAP model, chosen as a ``typical'' CS, was reddened over a range of $E(B-V)$ values using the reddening laws of \cite{cardelli1989} and of \cite{fitzptrick2007}, both with R$_{V}$=3.1. The colours of each model were calculated, and are listed in Table \ref{tab:synthetic_reddened_CS}. Colours using a $T_{\text{eff}}$=50kK, log(g)=7.0 model were also calculated, but the differences were found to be negligible.

\begin{figure}
	\begin{center}
		\includegraphics[clip, width=\linewidth]{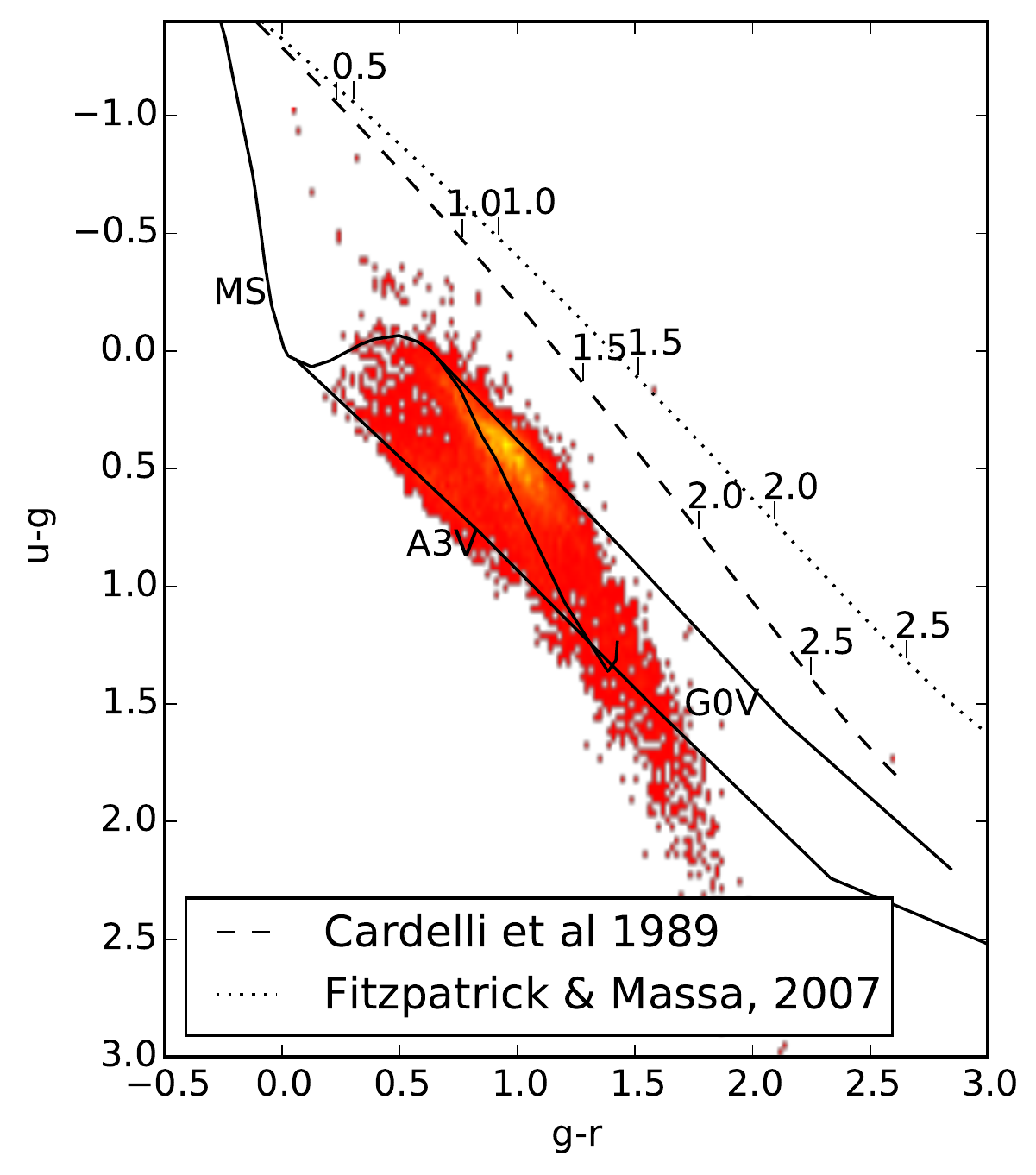} 
		\caption{A colour-colour plot using the calibrated, observed (not de-reddened) magnitudes of stars in VPHAS+ pointing 0757. Overlaid are the  main sequence (MS) locus, and the A3V and G0V reddening lines for $R=3.1$ from \protect\cite{drew2014}. The dashed lines show the CS reddening lines calculated using two reddening laws, with tick marks showing the values of $E(B-V)$.}
		\label{fig:cs_reddening_lines}
	\end{center}
\end{figure}

Figure \ref{fig:cs_reddening_lines} shows the ($u-g$) vs. ($g-r$) colour-colour plot of VPHAS+ pointing 1738, overlaid with the CS reddening lines derived from the two reddening laws, with tick marks indicating the values of $E(B-V)$. Also shown are the synthetic main sequence locus, and G0V and A3V reddening lines from \cite{drew2014}. As well as being used for calibrating the $u$ band, these plots allow true CS to be identified; we would expect the observed colours of any true CS to lie close to the predicted CS reddening line. For this purpose, the choice of reddening law will not impact our results, as it is clear a true CS will lie far from the main cluster of stars in the upper parts of the plot.

The second use of the predicted, reddened CS colours is to derive the value of $E(B-V)$ from an observed CS's $u-g$ colour, as described in Section \ref{sec:ir_excess_method}. From Table \ref{tab:synthetic_reddened_CS}, we can see that the difference in the $u-g$ colour predicted by each reddening law is very small, less than 0.04 for $E(B-V)<2.0$. There is a bigger impact on the $g-r$ colour, but that is not used in here. Therefore again, the choice of reddening law will have a negligible effect on our results, and we continue using the reddening law of \cite{cardelli1989}.

\subsubsection{Main sequence stellar colours}
\label{subsec:ms_predicted_colours}

In order to estimate the spectral type of a CS companion given an observed CS IR excess, the main sequence spectral flux library from \cite{pickles1998} was used to calculate the expected colours of main sequence stars in the VPHAS+ bands. These expected colours are presented in Table \ref{tab:synthetic_MS_colours}.

The expected absolute $V$ band magnitudes and $V-J$ colours from \cite{demarco2013}, combined with the colours calculated in this work were used to approximate the expected absolute magnitudes in the $J$ and $i$ bands ($M_{i}$ and $M_{J}$, respectively). 
These are later used to give an \emph{indication} of CS-companion spectral type.

\section{PN sample}
\label{sec:pn_sample}

There are approximately 1500 
currently known true or likely true PNe listed in the HASH database \citep{parker2016} within the footprint of VPHAS+. Around 830 of these can 

\begin{figure*}
	\begin{tabularx}{\linewidth}{@{}cXX@{}} 
		\begin{tabular}{ccc}
			
			\subfloat[Hf~38]{\includegraphics[width=5.5cm, height=4cm,keepaspectratio]{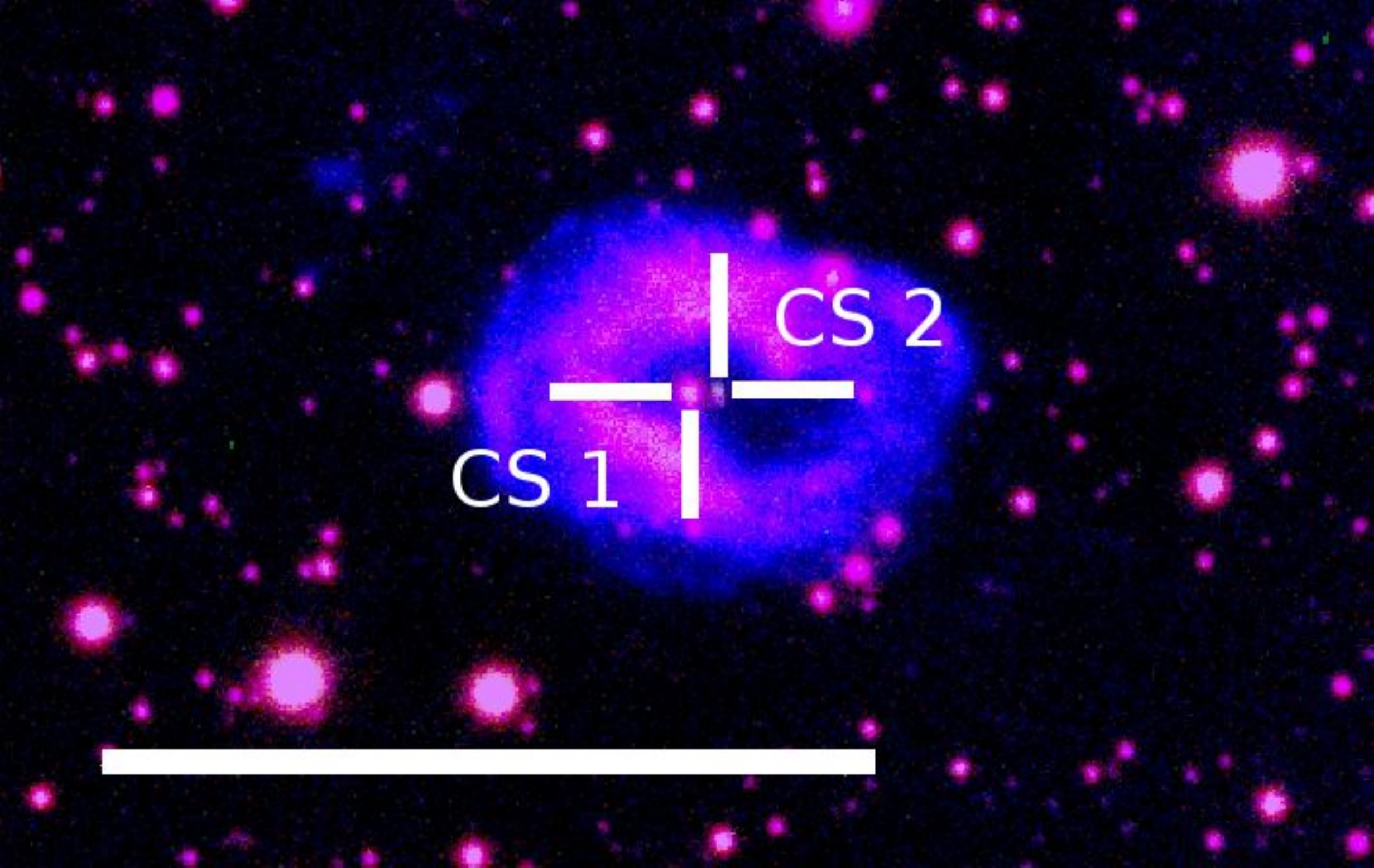}\label{fig:pn_sample_imgs_hf38}} &
			\subfloat[NGC~6337]{\includegraphics[width=5.5cm, height=4cm,keepaspectratio]{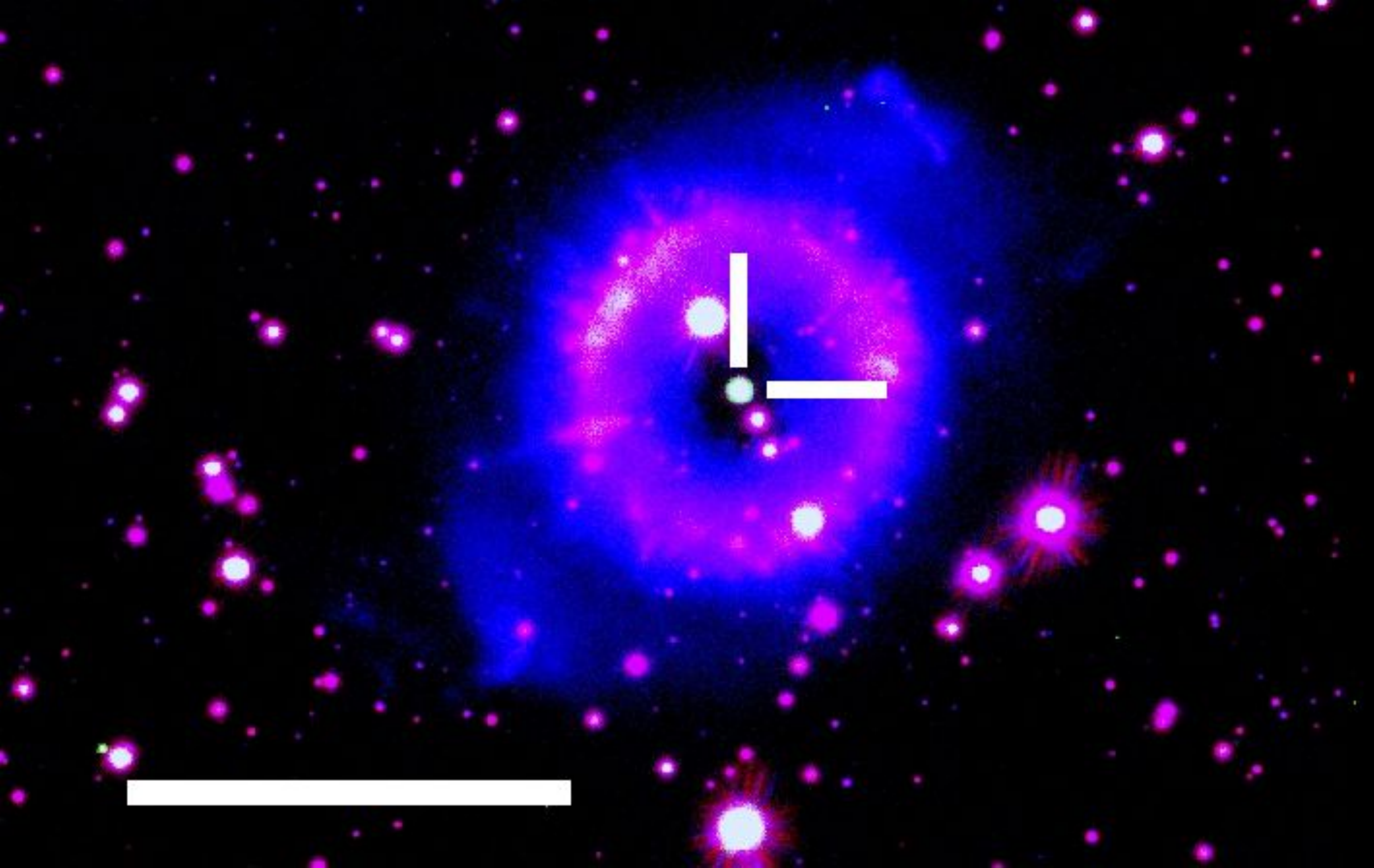}} &
			\subfloat[PNG~288.2+00.4]{\includegraphics[width=5.5cm, height=4cm,keepaspectratio]{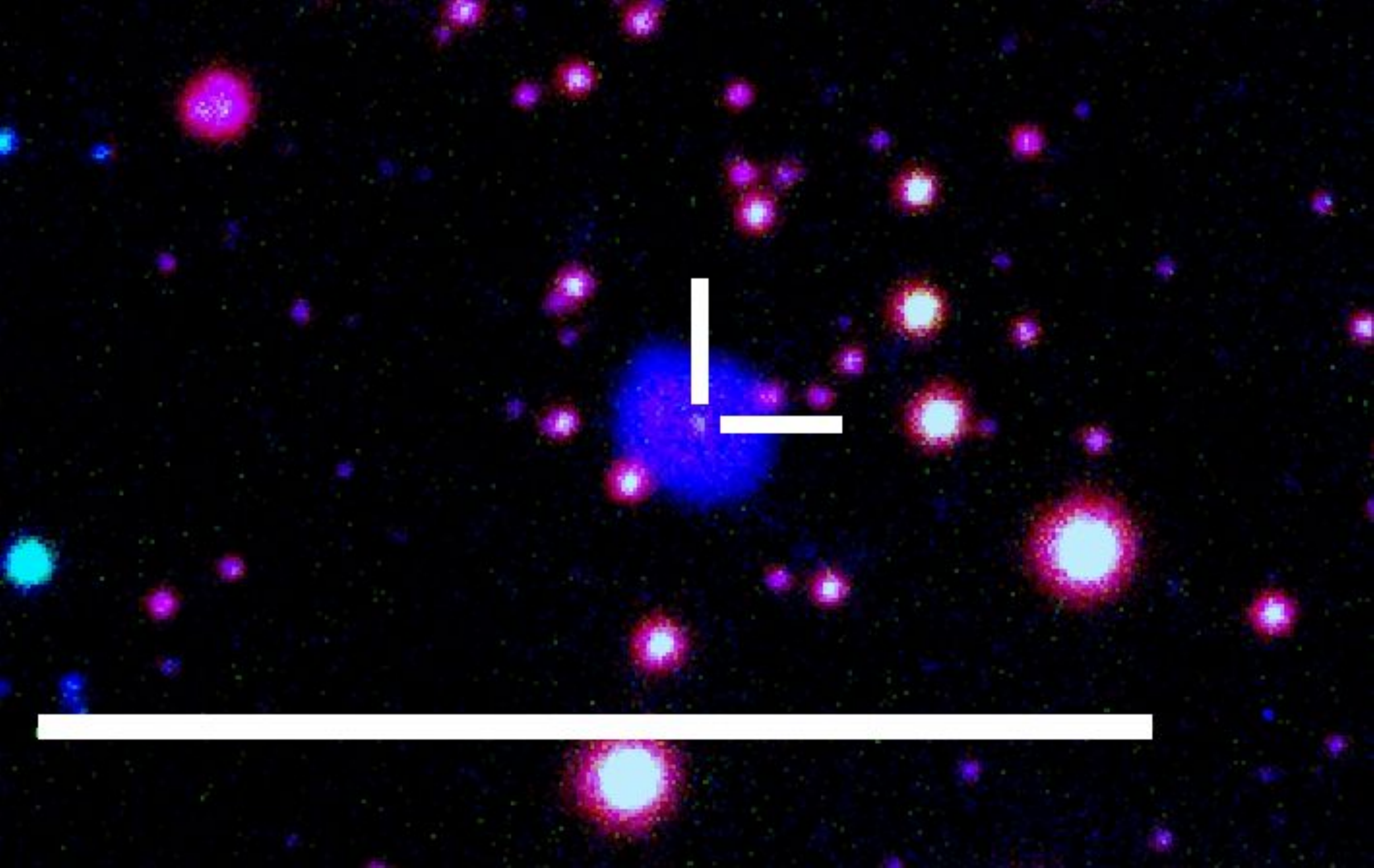}}\\

			\subfloat[PNG~003.4+01.4]{\includegraphics[width=5.5cm, height=4cm,keepaspectratio]{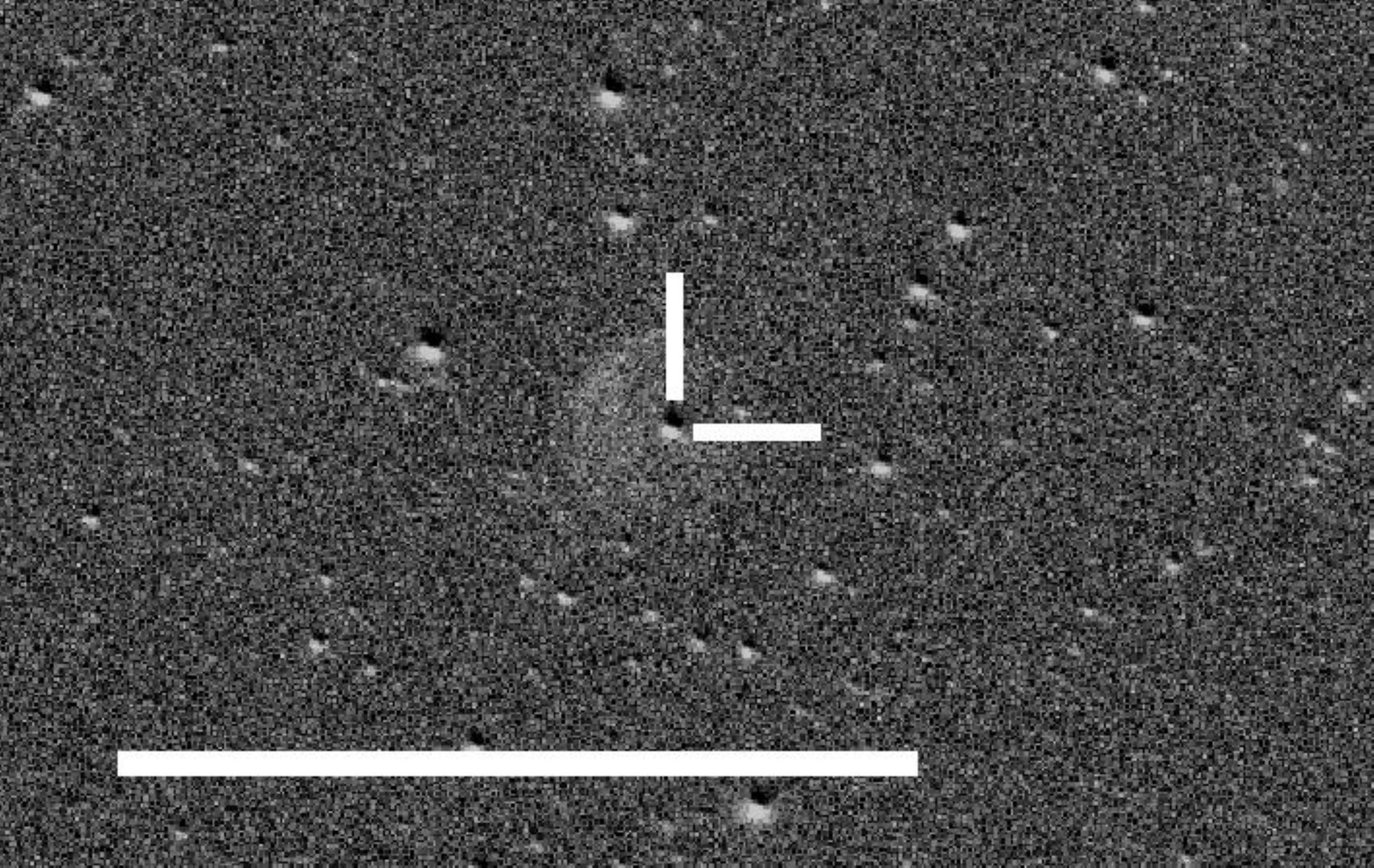}} &
			\subfloat[PNG~242.6-04.4]{\includegraphics[width=5.5cm, height=4cm,keepaspectratio]{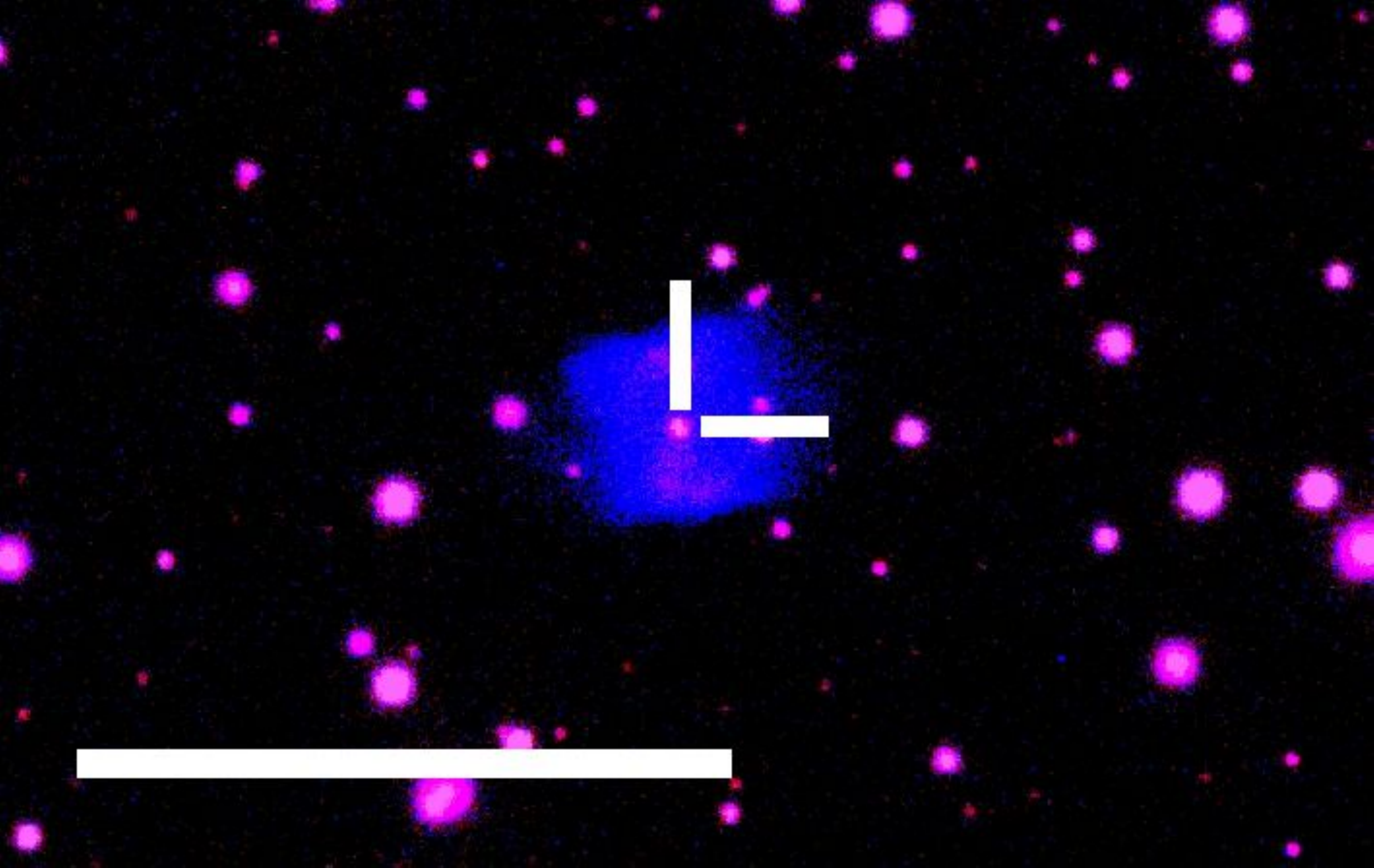}} &
			\subfloat[Pe~2-5]{\includegraphics[width=5.5cm, height=4cm,keepaspectratio]{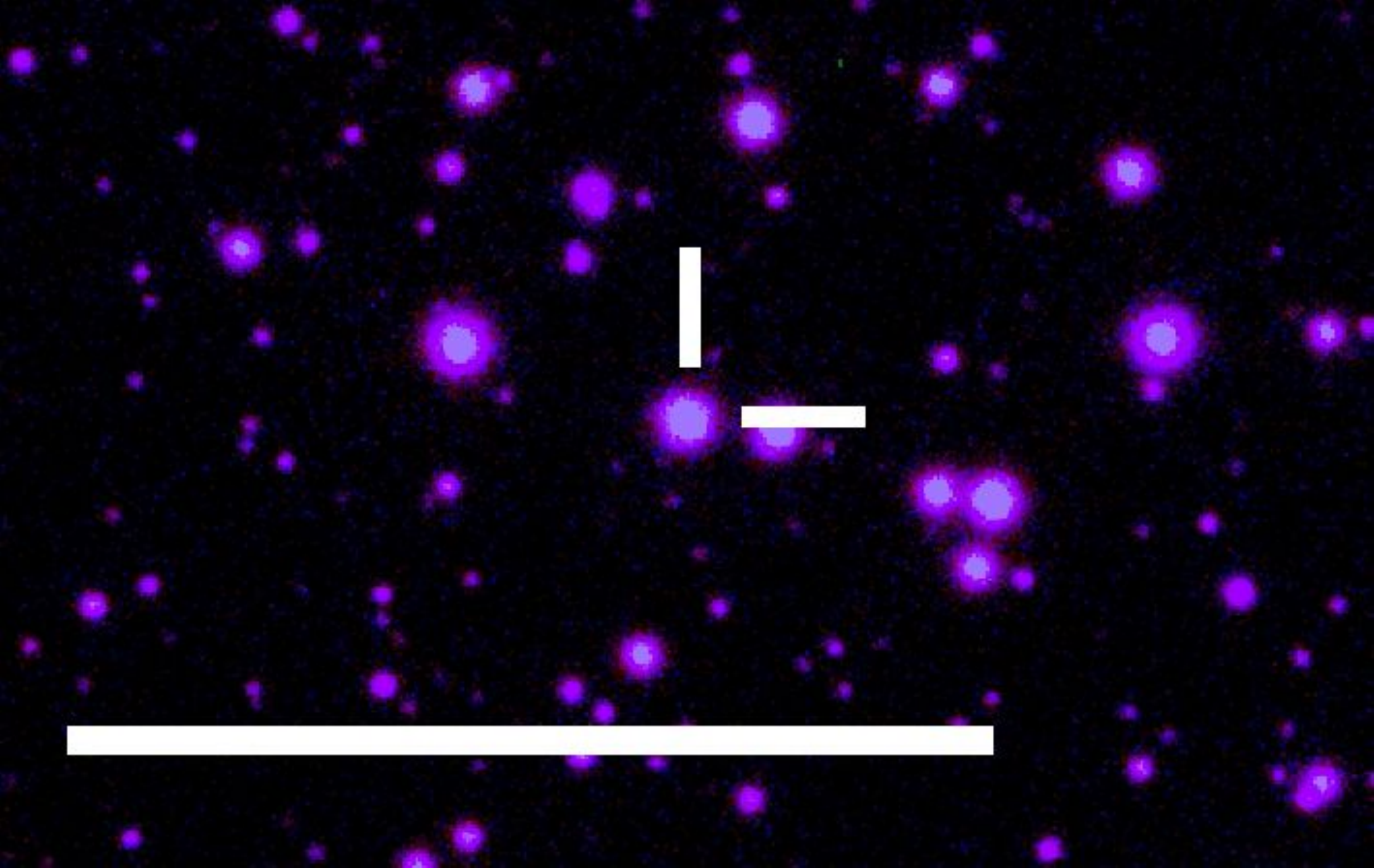}}\\
			
			\subfloat[PNG~293.4+00.1]{\includegraphics[width=5.5cm, height=4.0cm,keepaspectratio]{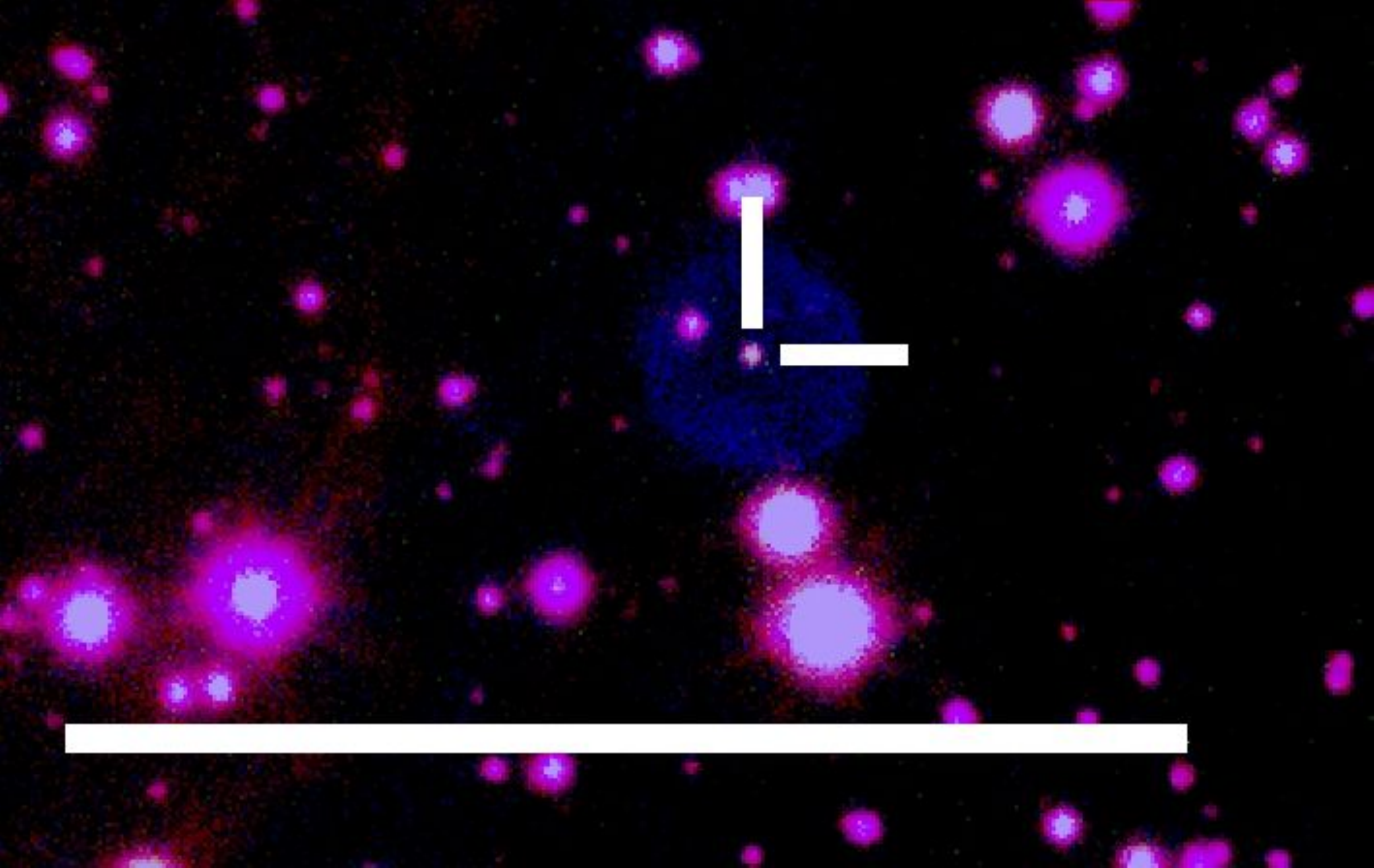}} &
			\subfloat[PNG~344.4+01.8]{\includegraphics[width=5.5cm, height=4cm,keepaspectratio]{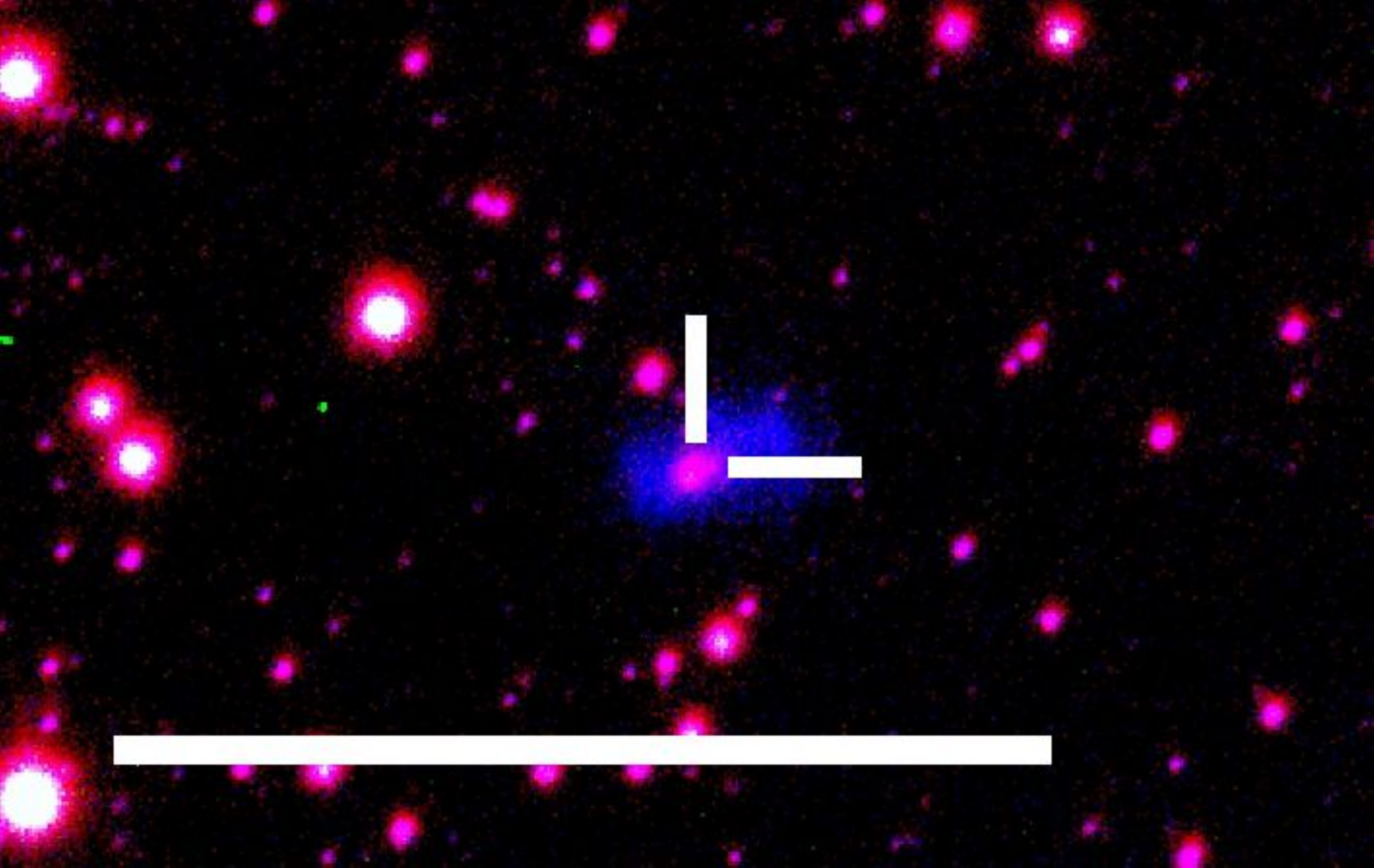}} &
			\subfloat[PNG~354.8+01.6]{\includegraphics[width=5.5cm, height=4cm,keepaspectratio]{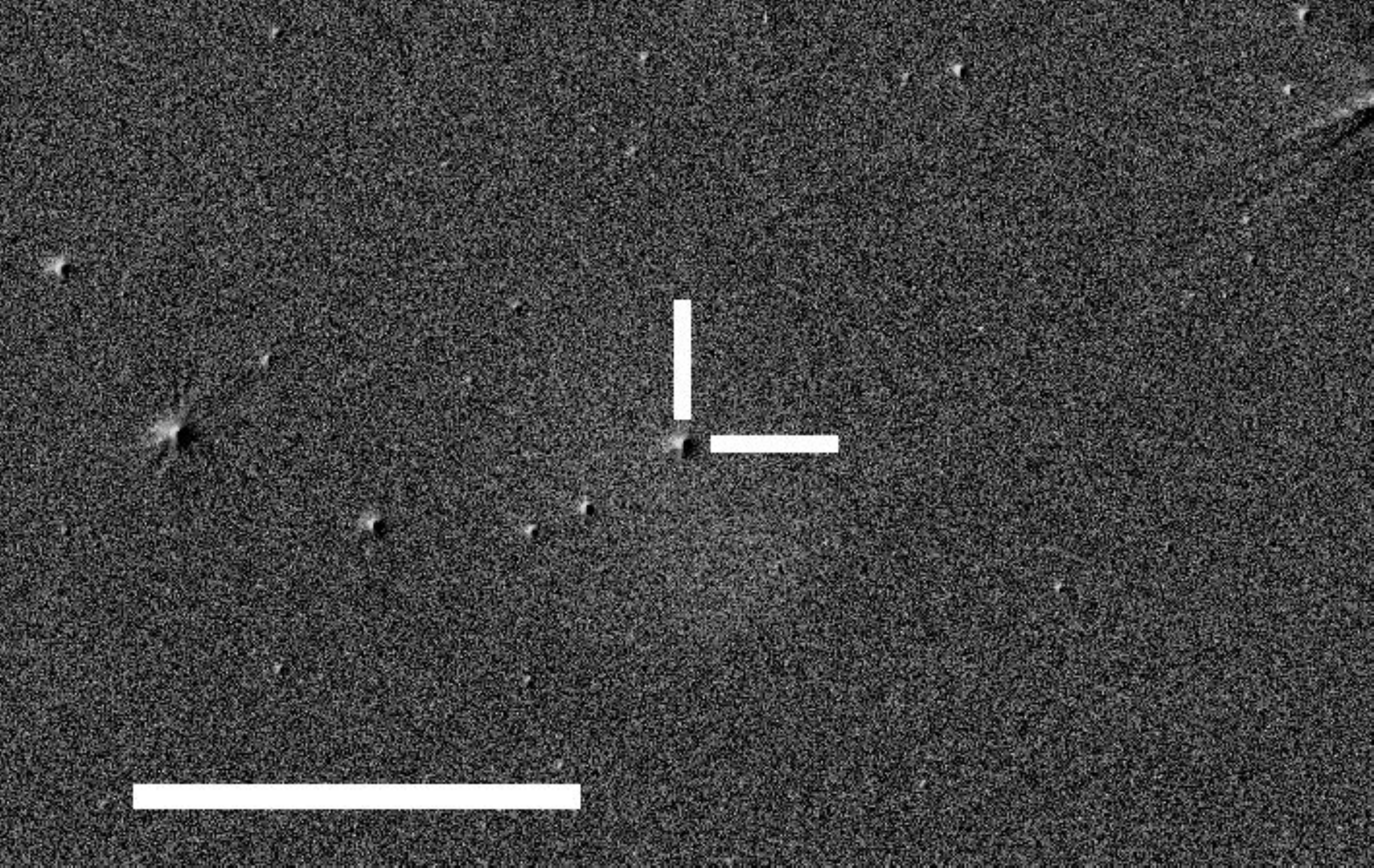}} \\
			
			\subfloat[PNG~355.9+00.7]{\includegraphics[width=5.5cm, height=4cm,keepaspectratio]{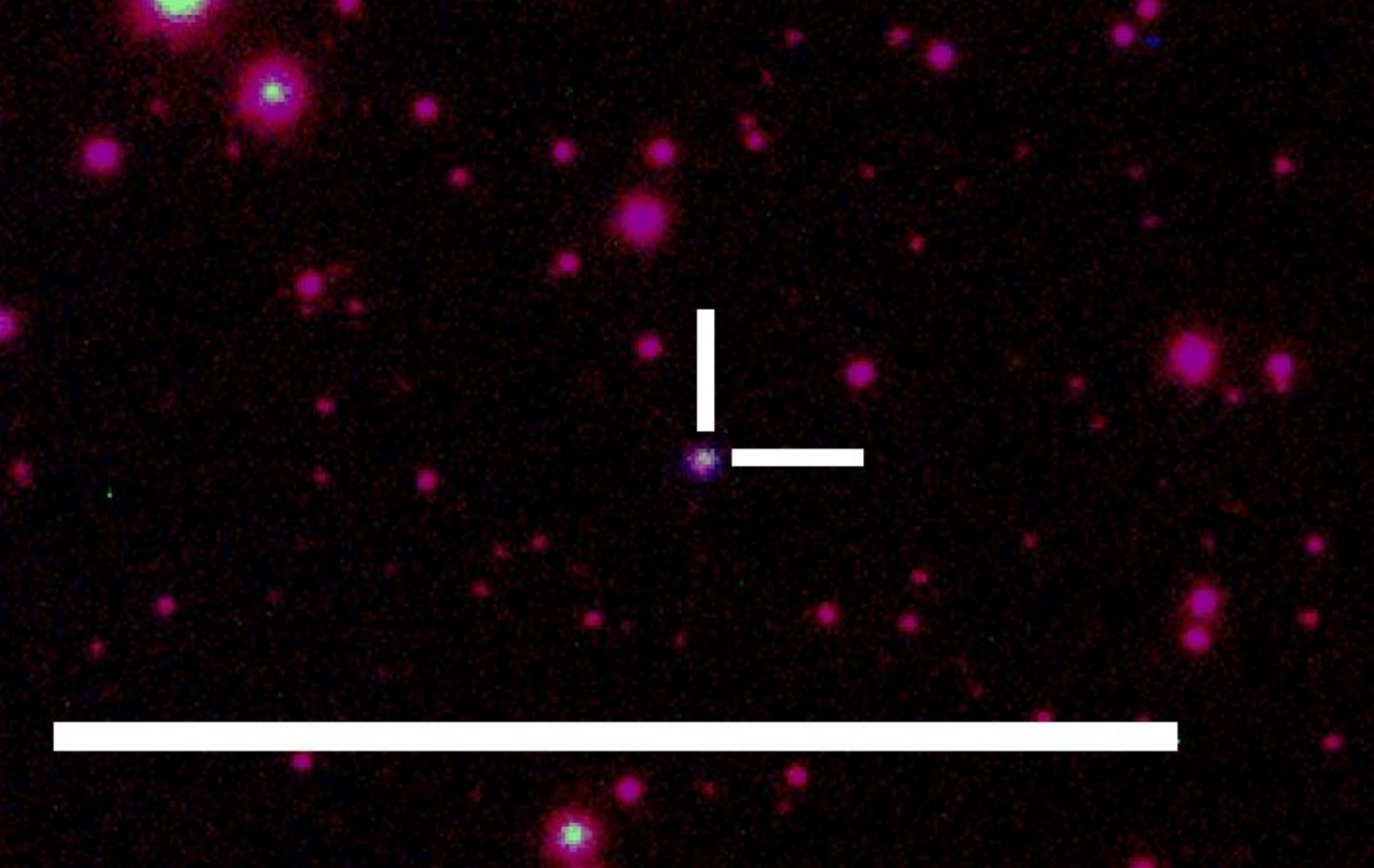}} &
			\subfloat[PTB~25]{\includegraphics[width=5.5cm, height=4cm,keepaspectratio]{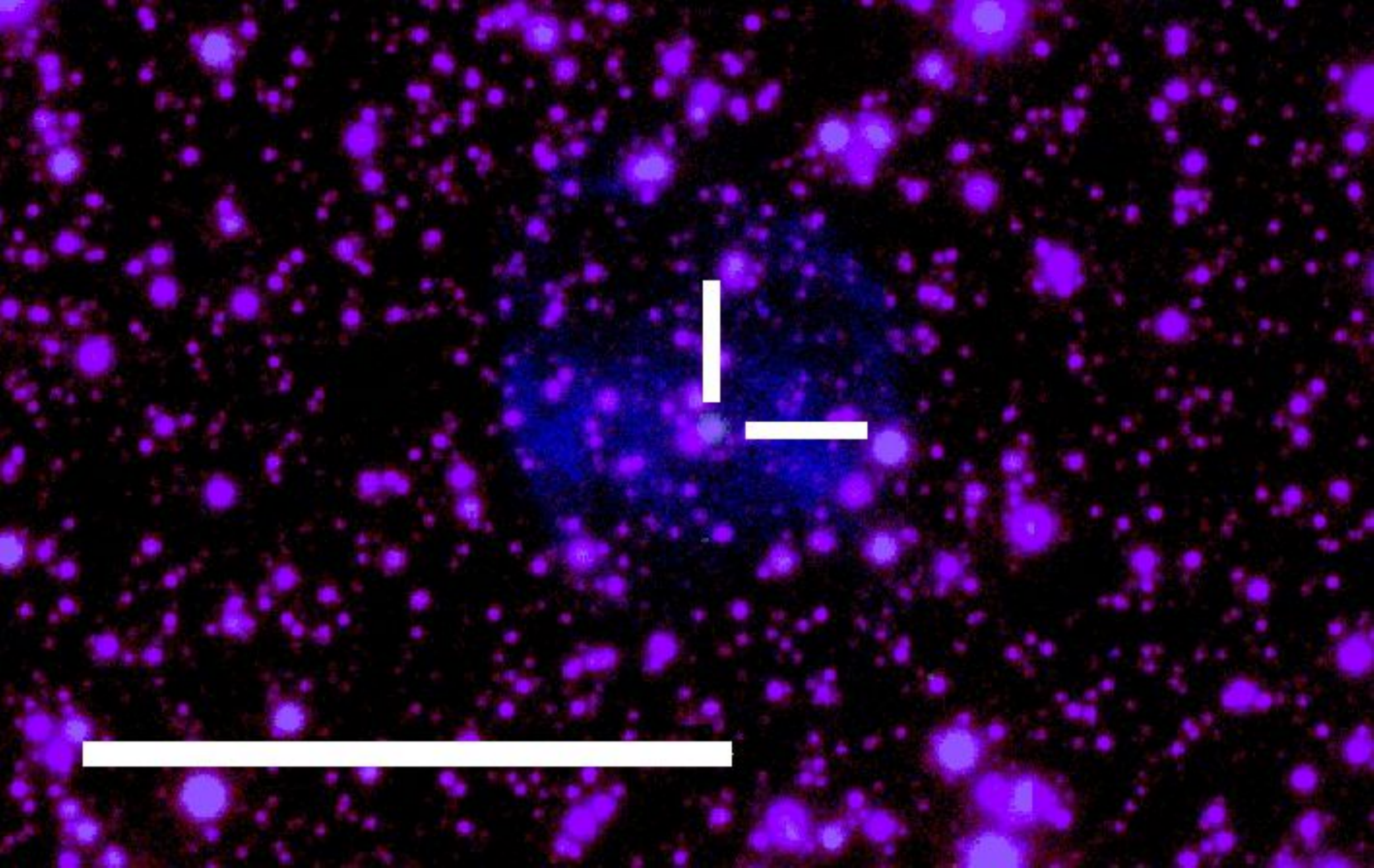}} &
			\subfloat[Abell~48]{\includegraphics[width=5.5cm, height=4.0cm,keepaspectratio]{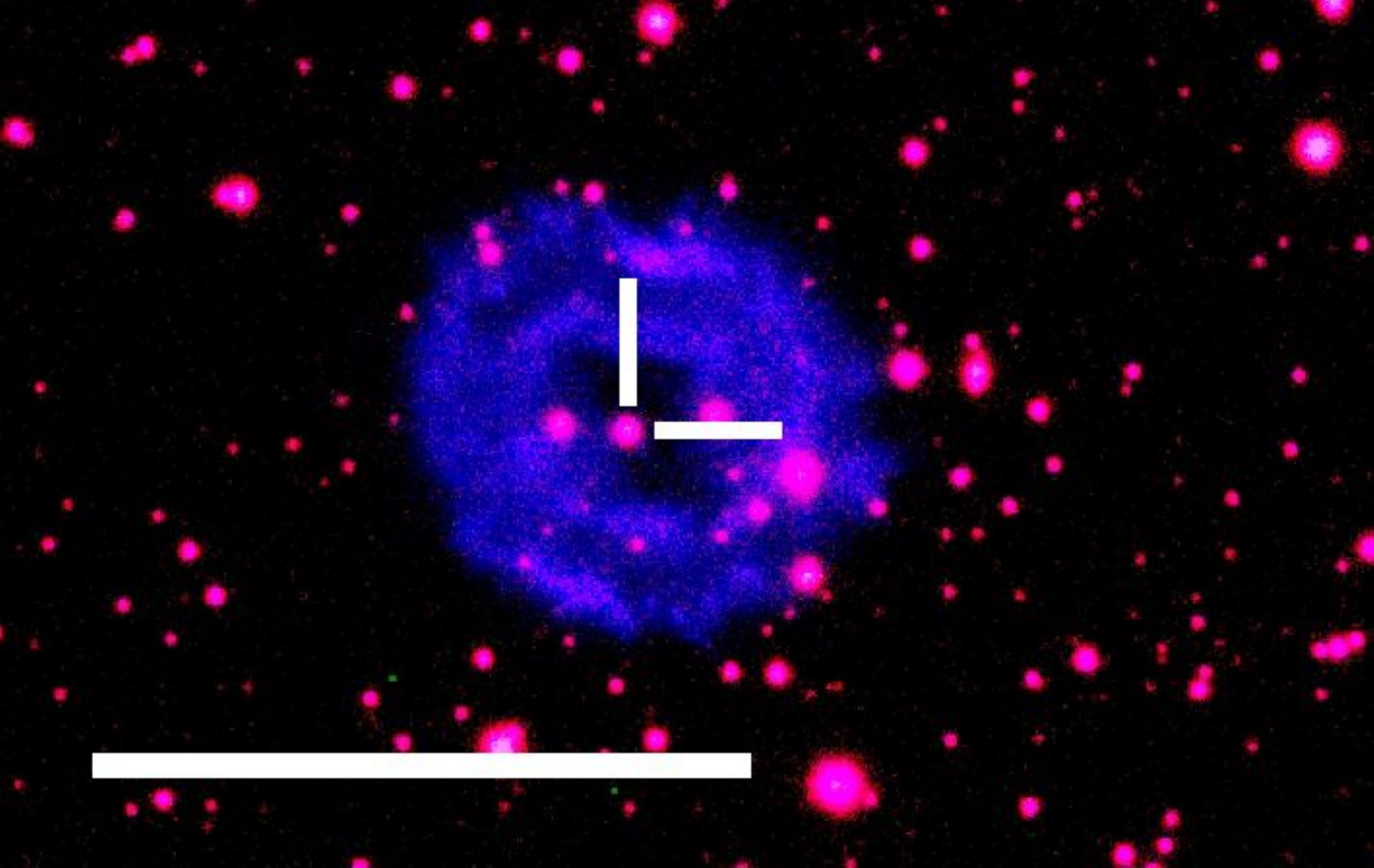}} \\
			
			&
			\subfloat[Sh~2-71]{\includegraphics[width=5.5cm, height=5.0cm,keepaspectratio]{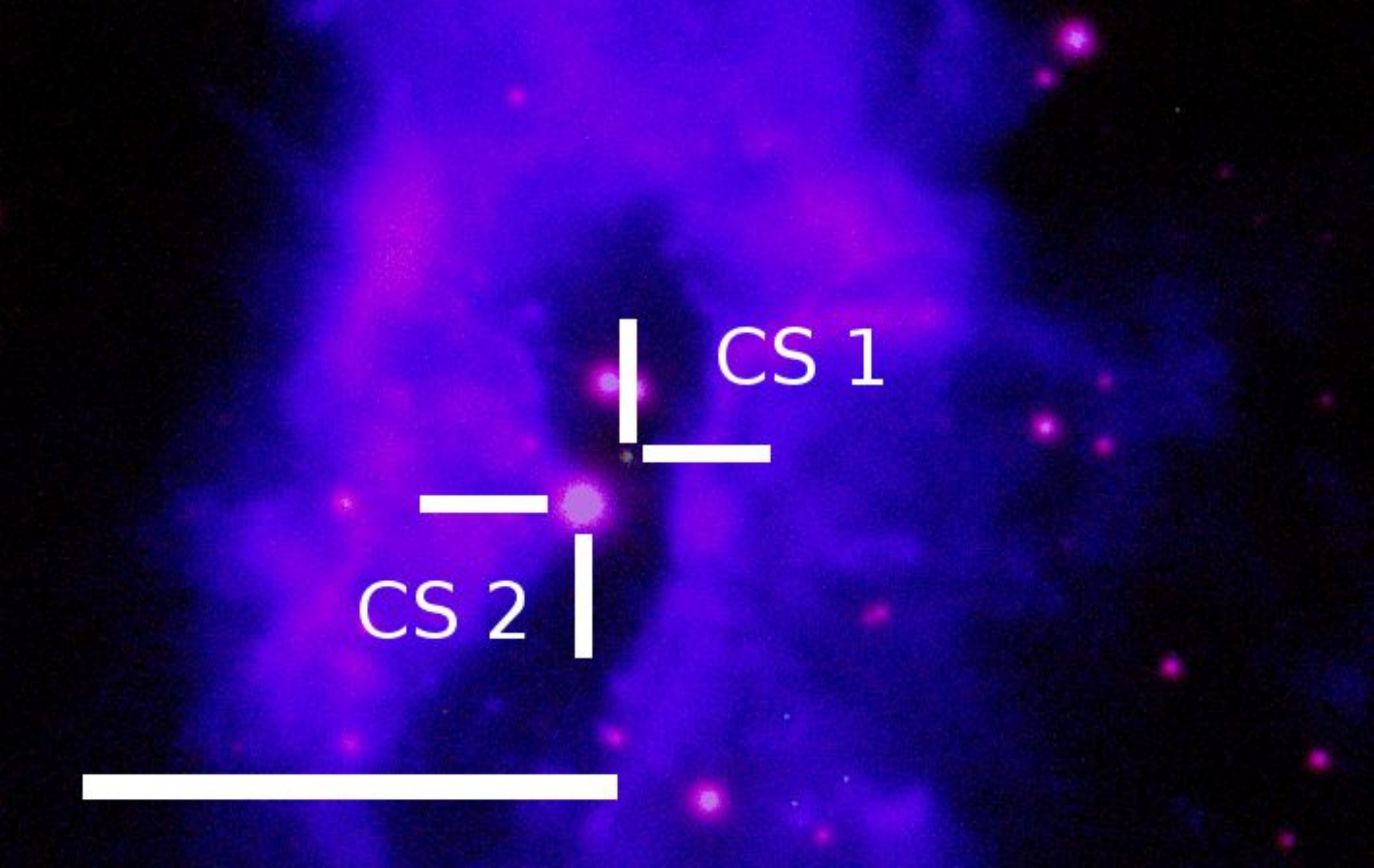}\label{fig:pn_sample_imgs_sh2-71}} &
			\\
			
		\end{tabular}
	\end{tabularx}
	\caption{VPHAS+ images of the PNe in our sample. Colour images show $r$, H$\alpha$ and $u$ band data in red, blue and green, respectively. Images in black and white are H$\alpha$/$r$ quotient images. The scale bar in the bottom left of each image indicates a length of 1 arcminute. }
	\label{fig:pn_sample_imgs}
\end{figure*}

\noindent
be found within the currently completed sections of the survey (up to DR3).

For this work, 90 true or likely true PNe within DR2 pointings fully competed before 30th September 2013 were examined to investigate whether VPHAS+ photometry is sufficiently precise enough to search for CS $i$ band excess. If successful, the techniques developed in this work can be applied to the remaining PNe within the VPHAS+ footprint Of the 90 PNe examined here, 20 have a known CS, or a visible candidate, that was detected in all the required $u$, $g$, and $i$ bands during the VPHAS+ catalogue generation process.

%
From this sample, the following CS were removed as the APASS data available was insufficient for calibration: PNG~000.6+03.1, PNG~001.5-02.8, PNG~002.3-01.7, PNG~002.9-03.0, PNG~018.8-01.6, PN~H~1-45 and PNG~001.2+02.8. Abell~48 was removed from our $i$ band excess investigation as it is a known Wolf-Rayet star (\citealt{todt2013}, \citealt{frew2014}) \footnote{Wolf-Rayet central stars constitute a sizeable fraction of all central stars. Their intrinsic colours are much redder and we cannot use our analysis on them.}, although its photometry is listed. Images of the remaining 12 PN and Abell~48 are shown in Figure \ref{fig:pn_sample_imgs}. The VPHAS+ catalogue and calibration details are summarised in Table \ref{tab:input_cat_data}, along with the seeing and calculated magnitude corrections for each observation, and the observation dates and times. There are two CS entries for Hf~38, labelled CS~1 and CS~2, as in this paper we suggest a new CS candidate. There are also two CS listed for Sh~2-71, again labelled CS~1 and CS~2, as the true CS is debated. The CS candidates for Hf~38 and Sh~2-71 are labelled in Figures \ref{fig:pn_sample_imgs_hf38} and \ref{fig:pn_sample_imgs_sh2-71} respectively.

\subsection{Removing central star impostors}
\label{subsec:cs_imposters}

\begin{figure}
	\begin{center}
		\includegraphics[clip, width=\linewidth]{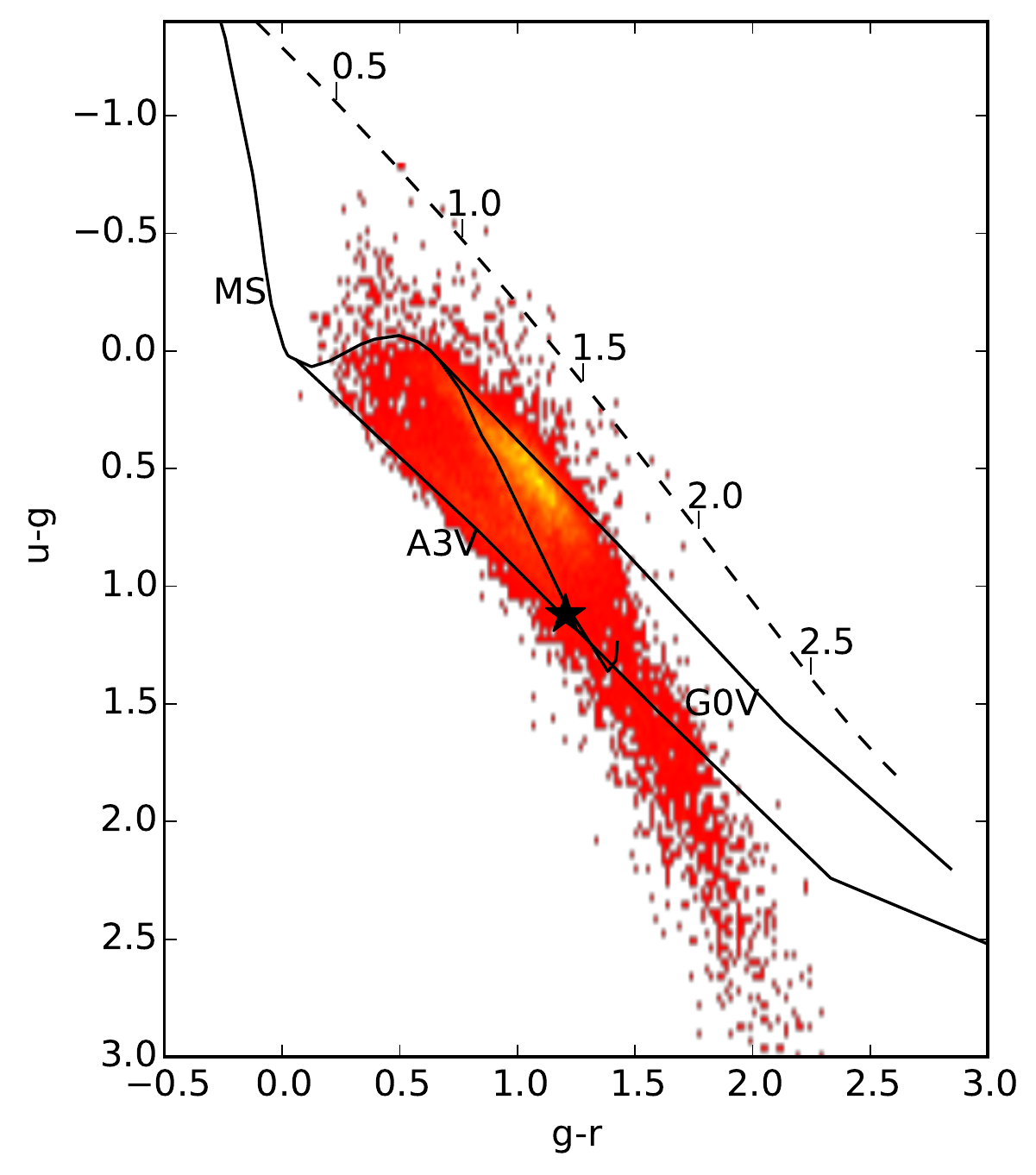} 
		\qquad
		\includegraphics[clip, width=\linewidth]{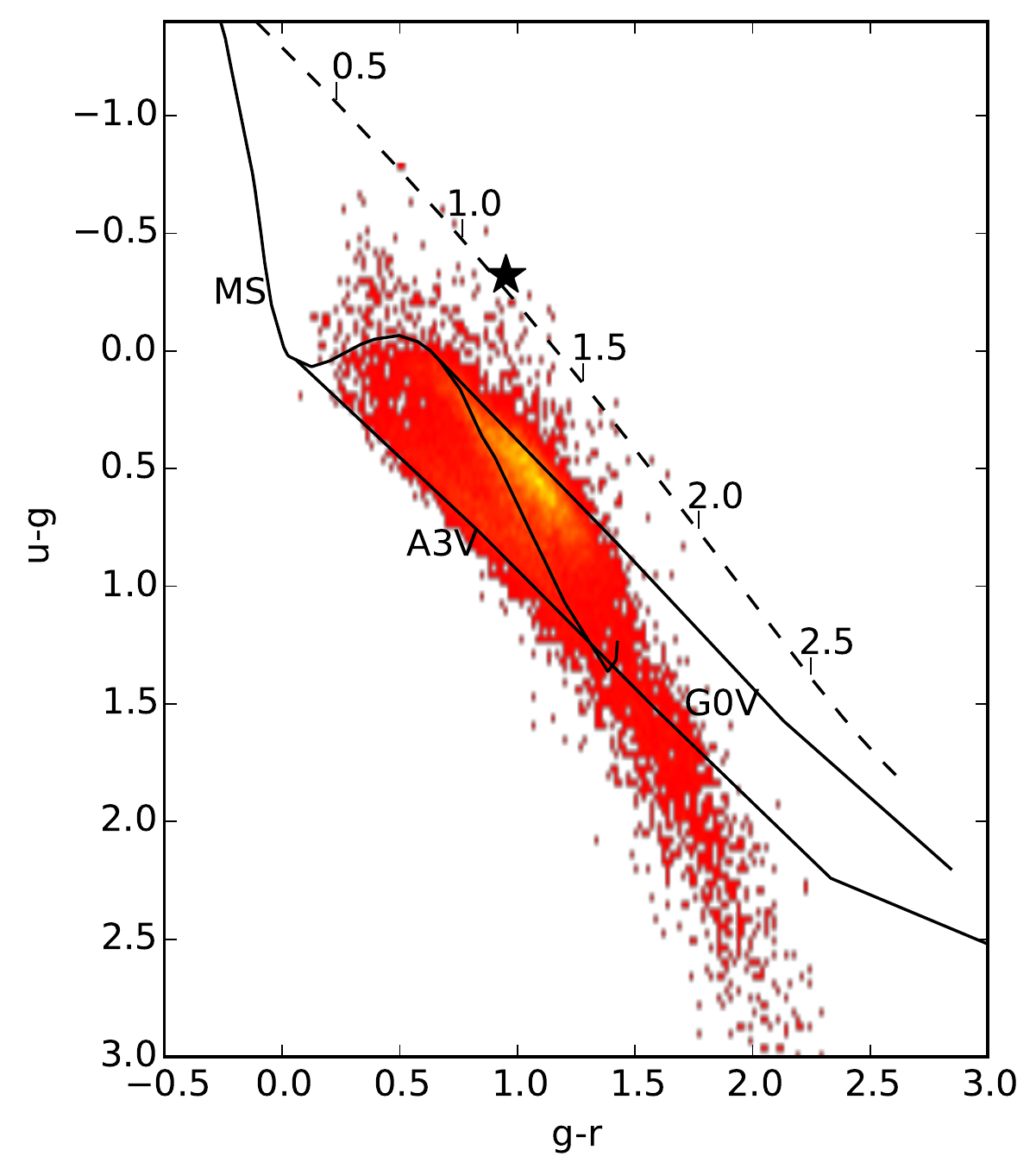}
		
		\caption{The observed colours of Hf~38~CS~1 (top) and Hf~38~CS~2 (bottom) indicated with a star, overlaid on the ($u-g$) vs. ($g-r$) colour-colour plot of objects in VPHAS+ pointing 1738 with main sequence (MS), G0V and A0 lines from \protect\cite{drew2014} and the CS reddening line using the reddening law of \protect\cite{cardelli1989}.}
		\label{fig:hf_38_colours}
	\end{center}
\end{figure}

The 14 CS candidates for the 12 PN in our sample were investigated to determine whether they are the true, or likely true CS. To be considered a true CS, the CS candidate had to be notably blue and well-centred in the PN. Table \ref{tab:final_pn_sample} summarises our findings and lists CS candidate coordinates in VPHAS+. A brief discussion of the rejected objects is provided below.

The CS candidate identified here for PNG~003.4+01.4 is the only star within the PN visible in VPHAS+. However, this star is not well centred in the nebula and a faint star closer to the geometric centre is visible in the VVV image. 

Upon first glance, PNG~354.8+01.6 appears to be a round PN in the VPHAS+ H$\alpha$/$r$ quotient image, centred on the star listed here. However, in the POPIPlaN \citep{boffin2012} H$\alpha$ image, the PN appears more elliptical with ISM interaction visible to the South-East of the nebula. Using this image, a new CS candidate was identified at Gaia coordinates 17:26:25.79 $-$32:22:01. Unfortunately, this object was not detected in VPHAS+.

Both PNG~355.9+00.7 and PTB 2-5 have been listed in the literature as PNe. However, no nebulae are visible in the VPHAS+ images of these objects, and neither have colours consistent with a CS. We therefore conclude that they are stars, as pointed out in the HASH database.

In previous images of Hf~38, only the brighter, westerly star (labelled CS~1 throughout this work) was visible. Here, we present a superior resolution image in which we are now able to identify a candidate CS located close to the centre of the nebula, labelled throughout this work as CS~2. Plotting the observed colours of both these objects on a ($u-g$) vs. ($g-r$) colour-colour plots, shown in Figure \ref{fig:hf_38_colours}, reveals that the colours of CS~2 lie close to the CS reddening line while those of CS~1 lies within the main group of stars. Hf~38~CS~1 was therefore rejected from our sample.

The CS candidate chosen for PNG~344.4+01.8 was the only star with photometry available in VPHAS+. However, it is off-centre and has colours inconsistent with those of a CS. Inspection of the VVV image suggests that our CS candidate is in fact two, possibly three, stars unresolved in VPHAS+. The true CS therefore remains unidentified.

\begin{table*}
	\begin{center}
		\caption{A summary of the objects kept, and those rejected from our CS sample as their location within the nebula and observed colours were inconsistent with a CS. Celestial coordinates are from VPHAS+. }
		\scalebox{0.8} 
		
		{\begin{tabular}{l c cc}
				\hline
				
				PN	& Common name & RA & DEC \\
				
				\hline
				\multicolumn{4}{l}{Possible CS}\\
				&&&\\
				
				PNG 018.8-01.9 & PTB 25 & 18:32:04.549 & -13:26:15.08 \\
				PNG 035.9-01.1 CS~1 & Sh~2-71 CS~1 & 19:01:59.954 & 02:09:16.17 \\
				PNG 035.9-01.1 CS~2 & Sh~2-71 CS~2 & 19:02:00.289 & 02:09:10.96 \\
				PNG 242.6-04.4 & - & 07:32:17.547 & -28:25:17.92 \\
				PNG 288.2+00.4 & - & 10:53:31.977 & -59:03:02.77 \\
				PNG 288.4+00.3 CS~2 & Hf~38 CS~2 & 10:54:35.170 & -59:09:46.67 \\
				PNG 293.4+00.1 & - & 11:30:58.948 & -61:15:50.96 \\
				PNG 349-01.1 & NGC~6337 & 17:22:15.673 & -38:29:01.67 \\

				\hline
				\multicolumn{4}{l}{Rejected objects}\\
				&&&\\
				
				PNG 003.4+01.4 & - & 17:48:15.558 & -25:15:14.34 \\
				PNG 285.4-01.1 & Pe 2-5 & 10:28:34.610 & -59:03:23.28 \\
				PNG 288.4+00.3 CS~1 & Hf~38 CS~1 & 10:54:35.456 & -59:07:46.92 \\
				PNG 344.4+01.8 & - & 16:54:43.326 & -40:41:47.05 \\
				PNG 354.8+01.6 & - & 17:26:25.801 & -32:21:49.84 \\
				PNG 355.9+00.7 & - & 17:32:44.493 & -32:01:07.44 \\

				\hline
		\end{tabular}}
		\label{tab:final_pn_sample}
	\end{center}
\end{table*}

\subsection{VPHAS+ photometry}
\label{subsec:sample_photometry}

Table \ref{tab:pn_input_mags} lists the photometry on the Vega system of the remaining 8 CS candidates in our sample, along with the best $J$ band magnitudes from the literature. The photometry of rejected objects can be found in Table \ref{tab:rejected_input_mags}. This photometry was obtained by calibrating each of the VPHAS+ pointings containing an object as described in Section \ref{subsec:vphas_calibration}. A suitable aperture for each object was chosen by considering the seeing in each filter and the amount of crowding around the CS.


Multiple measurements, available due to the VPHAS+ offset observing pattern, were combined provided our confidence in the calibration of each measurement did not differ significantly, and any changes in seeing were unlikely to increase nebula contamination of CS photometry. The C block $g$ band exposure, taken on a different date to the A block exposure, for Sh~2-71~CS~2 was not used, as this object is a known variable \cite{kohoutek1979}.


For the pointings in our sample, none of the red and blue observing blocks (containing the $r$ and $r2$ measurements, respectively), were completed on the same night. This allows us to examine the stability and consistency of VPHAS+ photometry, although it should be noted that the $r$ band magnitudes are not used in the IR excess method presented here. In general, we find that $r$ and $r2$ measurements agree to better than 0.027 magnitudes, except the $r$ and $r2$ measurements of Hf~38~CS~2 and PNG~288.2+00.4 (both in VPHAS+ pointing 1738) that differ by approximately 0.12 magnitudes. We find this discrepancy is consistent with all stars of a similar magnitude within the pointing, leading us to believe that it is due to a difference in the calculated calibration parameters, likely due to the large difference in seeing (listed in Table \ref{tab:input_cat_data}) between the two measurements. The calibration of the $r$ band is more consistent with APASS than that of $r$2, so is our preferred measurement.



\begin{table*}
	\begin{center}
		\caption{The calculated single-band VPHAS+ CS magnitude measurements on the Vega system. There are two $r$ band measurements as there is an exposure in the blue observing block (containing $u$, $g$ and $r2$ exposures), and an exposure in the red observing block (containing $r$, H$\alpha$ and $i$ exposures). The number of measurements is shown in brackets, and the average magnitude quoted where multiple measurements were available.}
		\scalebox{0.8}
		{\begin{tabular}{lc ccccc c}
				\hline
				PN & Common Name & $u$ & $g$ & $r2$ & $r$ & $i$ & $J$ \\
				\hline
				
				PNG 018.8-01.9 & PTB 25 & 17.296$\pm$0.026 (2) & 18.197$\pm$0.012 (3) & 17.683$\pm$0.013 (2) & 17.655$\pm$0.014 (2) & 17.227$\pm$0.020 (2) & - \\

				PNG 035.9-01.1 CS 1 & Sh~2-71 CS 1 & 19.499$\pm$0.055 (1) & 20.147$\pm$0.027 (2) & 20.195$\pm$0.047 (1) & 20.724$\pm$0.073 (1) & 19.082$\pm$0.033 (1) & - \\

				PNG 035.9-01.1 CS 2 & Sh~2-71 CS 2 & 14.918$\pm$0.053 (1) & 14.269$\pm$0.005* (1) & 13.287$\pm$0.007 (1) & 13.385$\pm$0.006 (1) & 12.887$\pm$0.007 (1) & 12.002$\pm$0.026$^1$ \\

				PNG 242.6-04.4 & - & 19.414$\pm$0.029 (1) & 18.926$\pm$0.012 (2) & 18.026$\pm$0.009 (1) & 18.012$\pm$0.008 (1) & 17.856$\pm$0.011 (1) & -\\
				
				PNG 288.2+00.4 & - & 20.055$\pm$0.041 (2) & 19.733$\pm$0.022 (3) & 18.672$\pm$0.015 (2) & 18.550 $\pm$0.016 (2) & 18.480$\pm$ 0.021 (2) & - \\
				
				PNG 288.4+00.3 CS 2 & Hf~38 CS 2 & 19.477$\pm$0.032 (1) & 19.794$\pm$0.023 (2) & 18.847$\pm$0.017 (1) & 18.726$\pm$0.022 (1) & 18.398$\pm$0.022 (1) & - \\	
			
				PNG 293.4+00.1 & - & 19.553$\pm$0.030 (1) & 20.021$\pm$0.027 (1) & 19.133$\pm$0.021 (1) & 19.094$\pm$0.017 (1) & 18.437$\pm$0.029 (1) & -\\
			
				PNG 349.3-01.1 & NGC 6337 & 15.026$\pm$0.032 (2) & 16.116$\pm$0.007 (3) & 15.790$\pm$0.007 (2) & 15.407$\pm$0.007 (2) & 15.238$\pm$ 0.010 (2) & 15.212$\pm$0.059$^1$\\

				\hline
				
				\multicolumn{7}{l}{$^1$ 2MASS } \\
				\multicolumn{7}{l}{* Multiple observations were not used as this is a known variable star.}	\\

		\end{tabular}}
		\label{tab:pn_input_mags}
	\end{center}
\end{table*}

\subsection{Distances}
\label{subsec:sample_distances}

The catalogue of \cite{frew2016} was searched to find distance measurements to each of the PN in the sample. These are summarised in Table \ref{tab:cs_luminosity}. For all other PN in the sample, distances were calculated using the same method.  It should be emphasised that the distance is \emph{not} used in the detection of IR excess, but is used to give an indication of companion spectral type to within a few subclasses.

\section{Analysis}
\label{sec:analysis}

\subsection{Interstellar reddening}
\label{subsec:analysis_reddening}

It is crucial that the $E(B-V)$ value determined using our method be uncontaminated and accurate, as explained in Section \ref{sec:ir_excess_method}. First, $E(B-V)$ values were calculated by comparing the observed CS $u-g$ colour to those in Table \ref{tab:synthetic_reddened_CS}, which lists the expected colours of a 100kK CS reddened by different $E(B-V)$ values when viewed in the VPHAS+ bands using the reddening law of \cite{cardelli1989}. The errors listed are the formal errors derived from the observed $u$ and $g$ colours. This will be the main source of uncertainty on the calculated $E(B-V)$ values, given the extremely weak dependence of the expected CS $u-g$ colour on temperature, as discussed in Section \ref{subsec:cspn_predicted_colours}. Several other methods to determine the reddening were then used to check the accuracy of this value, and are discussed below.

Spectra of the PN from \cite{acker1992} were downloaded from the HASH website where available, and the nebula Balmer decrement calculated, from the observed H$\alpha$/H$\beta$ line intensity ratios. Other spectra available via HASH were not used as they are not flux calibrated. The H$\alpha$ to H$\beta$ line ratio was computed by calculating the area under Gaussian curves fitted to the lines.

The literature was also searched to find calibrated H$\beta$ fluxes and radio fluxes of the PN and $E(B-V)$ calculated using the radio/H$\beta$ ratio. This is useful, as the extinction derived from the H$\alpha$/H$\beta$ ratio depends on the optical reddening law assumed, and this method does not (for a review of this subject see \cite{nataf2016}).

The reddening map of \cite{green2015} was searched using distances listed in Table \ref{tab:cs_luminosity}. While values from the reddening map are not accurate, large discrepancies from this value may indicate contamination of CS photometry.

The $E(B-V)$ values calculated for each CS in our sample and values from the literature are listed in Table \ref{tab:measured_E(B-V)}. A discussion of these values is provided below.

Similar to previous results, e.g., \cite{ruffle2004}, we find that extinction values derived from radio data are lower than those from optical data. Table \ref{tab:measured_E(B-V)} also makes it clear that radio $E(B-V)$ values for Sh~2-71 decrease with frequency. Therefore, only $E(B-V)$ values derived from optical data were compared to that derived from the observed $u-g$ colour.

The $E(B-V)$ value for Hf~38~CS~2 derived from the $u-g$ colour is larger than that derived by \cite{tylenda1992}, and that calculated in this work from the Balmer decrement using the spectrum from \cite{acker1992}. This is likely because of $g$ band contamination by Hf~38~CS~1, and possibly the fainter red star to the south of CS~1, visible in Figure 
\ref{fig:pn_sample_imgs_hf38}.

For NGC 6337, the colour $E(B-V)$ value is lower than that derived from the Balmer decrement calculated here using the spectrum from \cite{acker1992}, and by \cite{tylenda1992} using their own spectrum. This is unexpected, as any contamination would act to increase this value. However, there is good agreement between the colour $E(B-V)$ values and that calculated by Frew (priv. comm.), suggesting the reddening calculated from the nebula is not representative of the reddening of the CS. 

There is a large discrepancy between the line of sight reddening and the $E(B-V)$ value calculated for PNG~242.6-04.4, suggesting that the PN is highly self-reddened, or that there is extensive CS $g$ band contamination by a companion star.

For PTB 25, the $E(B-V)$ value derived from the $u-g$ colour is smaller than both values from \cite{boumis2003}, who calculated $E(B-V)$ from the Balmer decrement, and a second, larger value that accounted for higher extinction in the direction of the Galactic Bulge. The good agreement between the colour $E(B-V)$ value and that from Frew (priv. comm.) increases our confidence in the value. 

It is reasonable to assume that the $E(B-V)$ value of a PN derived from the CS colours will be similar to that derived from the nebula. Therefore, given the good agreement between the literature values and the $E(B-V)$ values derived from the $u-g$ colour of Sh~2-71 CS 1, it seems unlikely that CS~2 is the true CS. Sh~2-71~CS~2 is therefore removed from our sample.

\subsection{Confirming the central stars}
\label{subsec:analysis_cs}

To confirm that the objects in our sample are the true CS, the luminosity of the CS was estimated by scaling the 100kK and 150kK log(g)=7.0, solar abundance TMAP models of CS atmospheres to match the VPHAS+ de-reddened CS $g$ band magnitudes.

The $g$ band was used rather than the $u$ band due to the larger uncertainty of the $u$ band observations, and the $r$ and $i$ band were not used as they are more prone to contamination by low-mass stellar companions. De-reddened magnitudes were calculated using the adopted $E(B-V)$ value for the PN from Table \ref{tab:measured_E(B-V)} and the observed magnitudes from Table \ref{tab:pn_input_mags}. The total flux, $F$, of the scaled spectrum was then calculated, and the luminosity, $L$, obtained using the standard equation, $L= F / 4\pi d^2$, with distance values, $d$, from \cite{frew2016}, or calculated in a similar manner. 

The luminosities thus calculated, listed in Table \ref{tab:cs_luminosity}, are only approximate. All estimated luminosities except those of PNG~242.6-04.4 and PNG~288.2+00.4 are consistent with those of a typical CS \citep{miller2016}, suggesting the light in these two objects is dominated by that of a main sequence star.

\begin{table}
	\begin{center}
		\caption{Luminosity estimates of each of the CS in the sample using magnitudes de-reddened using the adopted $E(B-V)$ value and distance estimates from \protect\cite{frew2016}. Distances for three PNe were calculated in a similar manner. Luminosities were calculated assuming a CS temperature of 100 and 150kK. }
		\label{tab:cs_luminosity}
		\scalebox{0.8} 
		{\begin{tabular}{l cc cc}
				\hline
				PN & Distance [kpc] & \multicolumn{2}{c}{Luminosity [L$_\odot$]} \\
				&                & 100 kK & 150 kK \\ 
				\hline
				
				PTB 25         & 3.20$\pm$0.68 & 1300$\pm$560       & 4060$\pm$1740 \\
				Sh~2-71 CS 1   & 1.32$\pm$0.40 & 80$\pm$50          & 240$\pm$150  \\
				PNG 242.6-04.4 & 6.6$\pm$1.2   & 136 000$\pm$50 000 & 418 000$\pm$152 000  \\
				PNG 288.2+00.4 & 6.7$\pm$1.5   & 42 800$\pm$19 000  & 131 000$\pm$58 800  \\
				Hf~38 CS 2     & 2.25$\pm$0.74 & 820$\pm$540        & 2500$\pm$1650 \\ 
				PNG 293.4+00.1 & 7.7$\pm$1.4   & 4960$\pm$1800      & 15 200$\pm$5540  \\
				NGC 6337       & 1.45$\pm$0.43 & 1100$\pm$650       & 3300$\pm$1980 \\
				
				\hline
		\end{tabular}}
	\end{center}
\end{table}

\subsection{CS temperatures}
\label{subsec:sample_temps}

Temperatures were estimated for PNe with spectra and a total H$\beta$ flux observation in the literature using pyCloudy \citep{morisset2013}. These were calculated by assuming a spherically symmetric PN with typical abundances, and a grid of CS temperatures and luminosities from the cooling track parameters of \cite{vassiladiswood1994}. The models were reddened using the adopted $E(B-V)$ values for each PN, listed in Table \ref{tab:measured_E(B-V)}.

The model CS temperature that reproduced the literature H$\beta$ flux and H$\beta$/O[III] line ratios for each PN is listed in Table \ref{tab:literature_temperature}, along with CS temperature measurements derived from nebula observations found in the literature.

While not accurate, these pyCloudy models and a study of the spectra allows us to discount the lower stellar temperatures determined for NGC 6337 and Sh~2-71, and the high energy-balance temperature of Hf~38. HeII lines are observed in the spectra of Hf~38 and NGC 6337, meaning the CS temperature must be greater than $\sim$90kK. They are not observed in the spectrum of Sh~2-71, but that is a noisy spectrum, especially at shorter wavelengths, so it is likely the signal was simply too low. 

Based on these arguments, we assume the temperature shown in bold in Table \ref{tab:literature_temperature} for these three CS, and all other CS are assumed to have a temperature of 100kK. We can see from Table \ref{tab:synthetic_CS_colours} that the $u-g$ and $g-i$ colours of CS vary 
by $\sim$0.07 magnitudes over the 50kK - 140kK temperature range, meaning that we can be confident that this assumption will have a minimal effect on the detection of IR excess.

\section{Results}
\label{sec:results_excesses}

The de-reddened colours of each of the objects believed to be true CS were calculated and compared to the predicted colours of a single CS, listed in Table \ref{tab:synthetic_CS_colours}, to search for $i$ and $J$ band excesses. 

\begin{figure}
	\captionsetup[subfigure]{labelformat=empty}
	\centering
	
	\subfloat[]{
		\includegraphics[width=\linewidth]{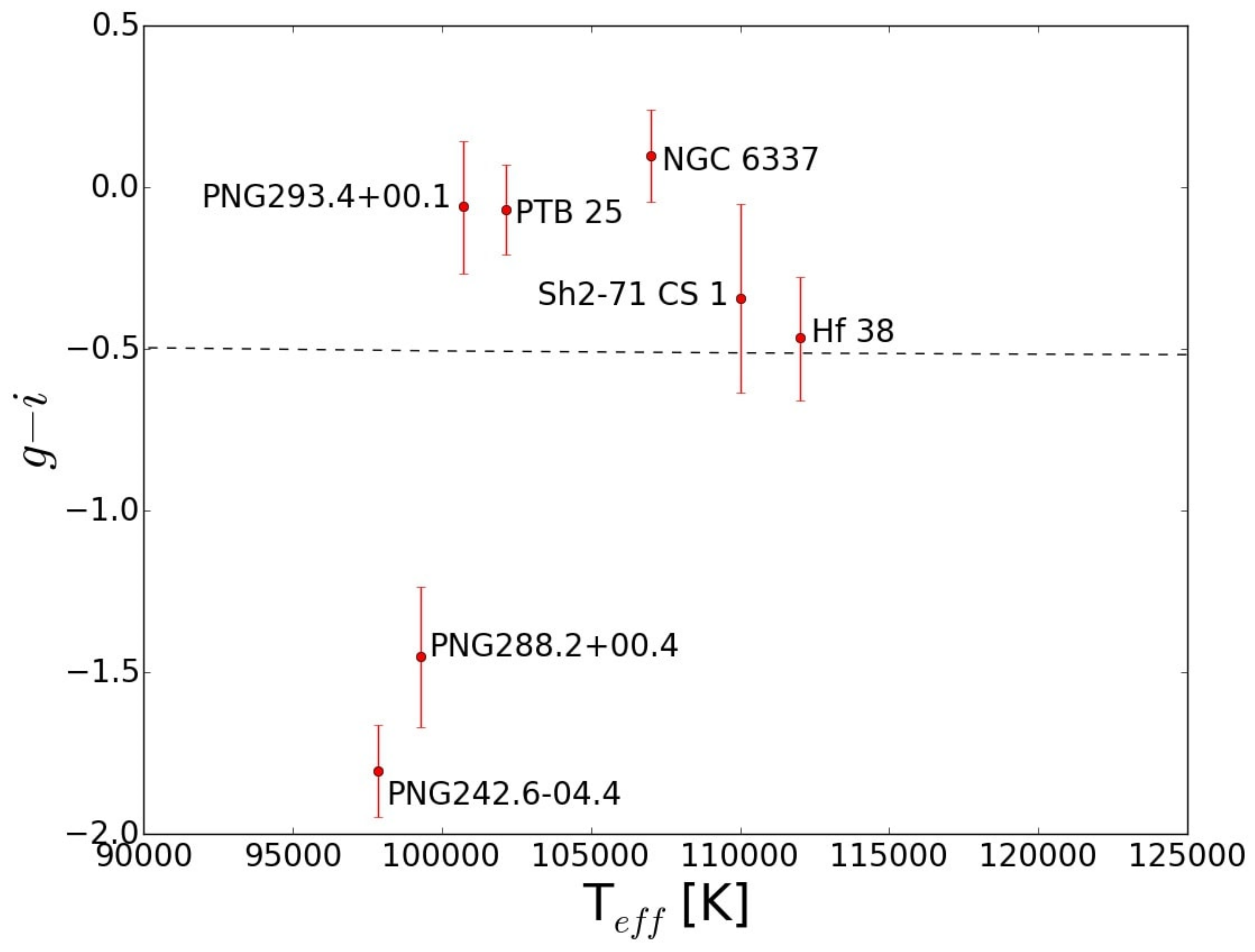}}
	
	\qquad
	
	\subfloat[]{
		\includegraphics[width=\linewidth]{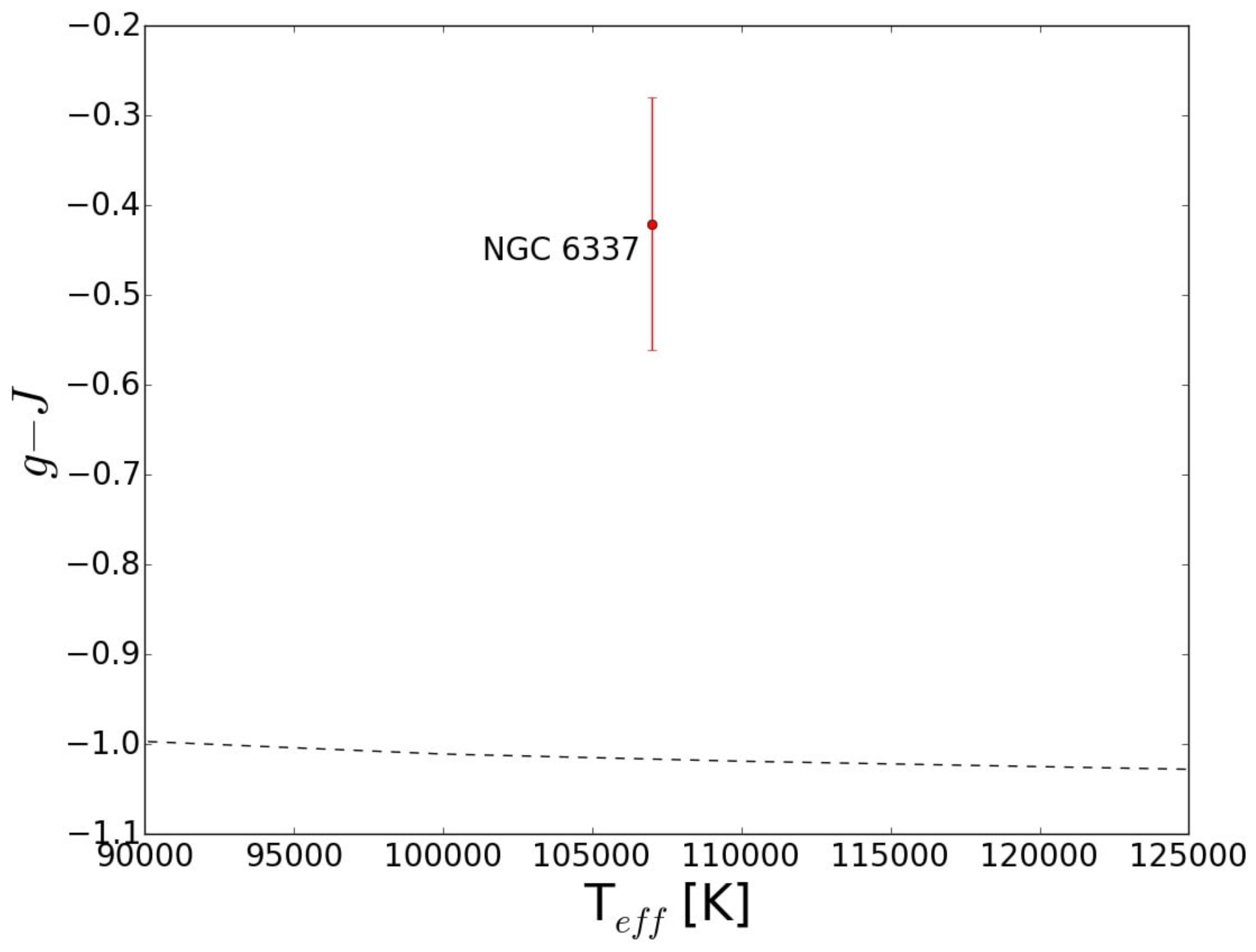}}

	\caption{The $g-i$ (top panel) and $g-J$ (bottom panel) colours of the CS in the sample, where photometry has been de-reddened using the adopted $E(B-V)$ value from Table \ref{tab:measured_E(B-V)}. The predicted colour of a single CS is shown as a black dashed line. Objects above the prediction line exhibit an $i$ or $J$ band excess consistent with the presence of a low mass main sequence companion. }
	\label{fig:excesses}
\end{figure}

Figure \ref{fig:excesses} shows the de-reddened $g-i$ and $g-J$ magnitudes and associated errors of the CS in the sample, de-reddened using the adopted $E(B-V)$ value listed in Table \ref{tab:measured_E(B-V)}. The black dashed lines show the predicted colour of a single star as a function of temperature. Log(g)=7.0 was assumed throughout and the CS assigned the same temperature have been shifted for clarity.  Objects whose $g-i$ (or $g-J$) colours are redder than that of a single star by more than can be justified by the uncertainty are binary candidates.  Objects below the line have an unphysical colour, suggesting the

\begin{landscape}
	\begin{table}
		\begin{center}
			\begin{threeparttable}
				\caption{$E(B-V)$ values calculated for each object using different methods. Reddenings from the reddening map of \protect\cite{green2015} use the distance listed in Table \ref{tab:cs_luminosity}.}
				\label{tab:measured_E(B-V)}
				\scalebox{0.8} 
				
				\begin{tabular}{l c cc ccc cc }
					\hline
					PN   & CS colour & \multicolumn{2}{c}{Balmer decrement} & \multicolumn{3}{c}{Radio/H$\beta$}           & Literature & Adopted \\ 
					name & $E(B-V)$    & Flux ratio          & $E(B-V)$         & $\log_{10}$F(H$\beta$) & Radio flux / frequency & $E(B-V)$ & $E(B-V)$ & $E(B-V)$ \\
					&           & H$\alpha$/H$\beta$ &                & [erg s$^{-1}$cm$^{-2}$]    & [mJy / GHz]  &        &        & \\
					\hline
					\multirow{3}{*}{Hf~38 CS 2} & \multirow{3}{*}{}{} & \multirow{3}{*}{7.237$^1$} & \multirow{3}{*}{0.91} & \multirow{3}{*}{-12.3$^1$} & \multirow{3}{*}{39$^2$ / 0.843} & \multirow{3}{*}{0.91} & 0.82$^{13}$ & \multirow{3}{*}{1.14$\pm$0.05} \\
					& 1.14$\pm$0.05 & & & & & & 1.11$^{14}$ & \\                                   
					& & & & & & & 1.13$^{15}$ & \\
					\multicolumn{9}{l}{}\\
					\multirow{2}{*}{NGC 6337} & \multirow{2}{*}{0.48$\pm$0.03} & \multirow{2}{*}{6.295$^1$} & \multirow{2}{*}{0.78} & \multirow{2}{*}{-11.35$^3$}  & 105.7$^2$ / 0.843 & 0.54 & 0.69$^{13}$ & \multirow{3}{*}{0.48$\pm$0.03} \\ 
					&&&&& 103$^4$ / 5.0 & 0.56 & 0.66$^{14}$ & \\ 
					&&&&& 126$^5$ / 14.7 & 0.68 & 0.41$^{19}$ & \\   
					
					\multicolumn{9}{l}{}\\
					\multirow{2}{*}{PNG242.6-04.4} & \multirow{2}{*}{1.77$\pm$0.03} & \multirow{2}{*}{-} & \multirow{2}{*}{-} & \multirow{2}{*}{-} & \multirow{2}{*}{-} & \multirow{2}{*}{-} & 0.47$\pm0.037$$^{15}$ &  \multirow{2}{*}{1.77$\pm$0.03 *} \\
					&&&&&&& 0.51 $^{19}$ & \\
					\multicolumn{9}{l}{}\\
					PNG288.2+00.4 & 1.65$\pm$0.05 & - & - & - & - & - & - & 1.65$\pm$0.05 *\\
					\multicolumn{9}{l}{}\\
					PNG293.4+00.1 & 1.00$\pm$0.05 & - & - & - & - & - & - & 1.00$\pm0.05$ \\ 
					\multicolumn{9}{l}{}\\
					\multirow{4}{*}{PTB 25} & \multirow{4}{*}{0.64$\pm$0.03} & \multirow{4}{*}{-} & \multirow{4}{*}{-} & \multirow{4}{*}{-} & \multirow{4}{*}{5.9$^7$$^8$$^9$ / 1.4}&  & 0.77$\pm0.0375$$^{15}$ & \multirow{4}{*}{0.64$\pm$0.03}\\ 
					& & & & & & &  0.69 $^{16}$ & \\
					& & & & & & & 1.44 $^{16}$ & \\ 
					& & & & & & &  0.43$\pm$0.14 $^{18}$ & \\
					\multicolumn{9}{l}{}\\
					\multirow{2}{*}{Sh~2-71 CS 1} & \multirow{2}{*}{0.86$\pm$0.76}  & \multirow{6}{*}{ - } & \multirow{6}{*}{ - } & \multirow{6}{*}{-11.6$^{10}$} & 57.9$^9$ /1.4 & 0.56 & 0.993$\pm$0.143$^{10}$ & \multirow{6}{*}{ 0.86$\pm$0.07 } \\
					& & & & & 63$^{12}$ / 1.5 & 0.59 & 0.83$\pm0.0555$$^{15}$ &  \\ 
					
					\multirow{3}{*}{Sh~2-71 CS 2} & \multirow{4}{*}{1.91$\pm$0.05} & & & & 78$^{11}$ / 4.85 & 0.69 & 0.81$\pm$0.13$^{17}$ & \\  
					
					&                                         & & & & 83$^{4}$ / 5.0 & 0.71 & 1.13 $^{1}$ & \\
					&                       & & & & 119$^5$ / 14.7 & 0.85 &  &\\
					
					\bottomrule
					
				\end{tabular}
				
				\begin{tablenotes}
					\small
					\item $^1$ \cite{acker1992}, 
					$^2$ \cite{murphy2007}, 
					$^3$ \cite{perek1971}, 
					$^4$ \cite{milne1975}, 
					$^5$ \cite{milne1982}, 
					$^6$ \cite{parker2006}, 
					$^7$ \cite{bojicic2011}, 
					$^8$ \cite{luo2005}, 
					$^9$ \cite{condon1998}, 
					$^{10}$ \cite{kaler1983}, 
					$^{11}$ \cite{becker1991}, 
					$^{12}$ \cite{zijlstra1989}, 
					$^{13}$ \cite{tylenda1992}, 
					$^{14}$ \protect{\cite{schlafly&finkbeiner2010}}, 
					$^{15}$ \cite{green2015}, 
					$^{16}$ \cite{boumis2003}, 
					$^{17}$ \cite{giammanco2011}, 
					$^{18}$ \cite{frew2016},
					$^{19}$ Frew, priv. comm.,
					* Not believed to be the true value, see Section \ref{sec:individual_pn}.
				\end{tablenotes}
			\end{threeparttable}
		\end{center}
	\end{table}
\end{landscape}

\begin{table}
	\begin{center}
		\begin{threeparttable}
			\caption{Temperatures from the literature for 3 of the sample objects, and Cloudy temperatures calculated using literature H$\beta$/O[III] line ratios from \protect\cite{acker1992}, and H$\beta$ fluxes from various sources, indicated in superscript. The temperature adopted in this work is shown in bold. For the remaining objects, we have no temperature estimates and a temperature of 100kK is assumed.}
			\label{tab:literature_temperature}
			\scalebox{0.8}
			
			\begin{tabular}{l cc c}
				\hline
				PN & \multicolumn{2}{c}{pyCloudy Temp.} & Literature \\
				& $\log_{10}$F(H$\beta$ )[K] & H$\beta$/O[III] [K] & Temp. [K] \\
				\hline
				&&&\\
				Hf~38 & \textbf{112 000}$^8$ & 115 000 & 236 000$^1$ \\
				&&&\\
				\multirow{5}{*}{NGC 6337} & \multirow{5}{*}{\textbf{107 000}$^9$} & \multirow{5}{*}{123 000} & 41 700$^2$ \\ 
				&&& 44 870$^3$ \\
				&&& 44 000$\pm$10 600$^4$ \\ 
				&&& 105 000 $^5$ \\ 
				&&& $>$90 000$^6$ \\                         
				&&& \\
				Sh~2-71 & \textbf{112 000}$^{10}$ & 115 000 & 29 500$\pm$10 200$^7$ \\                    
				\hline
			\end{tabular}	
			
			\begin{tablenotes}
				\small
				\item $^1$ \cite{preite-martinez1991}, $^2$ \cite{gorny1997}, $^3$ \cite{stanghellini1993}, $^4$ \cite{shaw1989}, $^5$ \cite{amnuel1985}, $ ^6$ \cite{hillwig2006}  $^7$ \cite{stasinska1997}, $^8$ \cite{acker1992}, $^9$ \cite{perek1971},  $^{10}$ \cite{kaler1983}
			\end{tablenotes}
		\end{threeparttable}
	\end{center}
\end{table}

\noindent
object is not a true CS, $E(B-V)$ has been over-estimated due to contamination, or a brighter companion star is dominating the light.

The $E(B-V)$ values, $i$ band excess and $J$ band excess, and an estimate of companion spectral type based on each excess for each PN in our sample are shown in Table \ref{tab:excess_results}. Companion spectral types were estimated by comparing the observed $g-i$ and $g-J$ excess to the predicted colours of main sequence stars in Table \ref{tab:synthetic_MS_colours}. The algorithm used preferentially chose the later spectral type in cases with a degenerate solution, as from Figure \ref{fig:contour_plot} we know this detection method can only detect spectral types later than G0V.

\begin{table*}
	\begin{center}
		\caption{$i$ and $J$ band excesses of the CS sample, and an estimate of the spectral type of the companion needed to produce this excess. Upper and lower limits are shown in square brackets. CS magnitudes were de-reddened using the adopted $E(B-V)$ values. }
		\scalebox{0.8} 
		
		{\begin{tabular}{l c cccc}
				\hline
				
				PN Name	&	$E(B-V)$	& $(g-i)_{0}$	& $\Delta(g-i)$	& M$_{i,2}$ &	Comp. spec. type\\ 		
				\hline
				
				Hf~38~CS~2 & 1.14$\pm$0.05 & -0.465$\pm$0.191 & 0.047$\pm$0.191 & $>$6.456 & later than K7V \\ 
				NGC~6337 & 0.48$\pm$0.03 & 0.097$\pm$0.142 & 0.609$\pm$0.142 & 5.968 [ 6.255 - 5.681 ] & K5V [ K5V - K3V ] \\ 
				PNG242.6-04.4 & 1.77$\pm$0.03 & -1.805$\pm$0.143 & -1.299$\pm$0.143 & - & - \\ 
				PNG288.2+00.4 & 1.65$\pm$0.05 &-1.45$\pm$0.217 & -0.944$\pm$0.217 & - & - \\ 
				PNG293.4+00.1 & 1.00$\pm$0.05 & -0.06$\pm$0.204 & 0.446$\pm$0.204 & 8.909 [ 9.321 - 8.496 ] & G5V [ K0V - G2V] \\ 
				PTB 25 & 0.64$\pm$0.03 & -0.069$\pm$0.14 & 0.437$\pm$0.14 & 5.919 [ 6.201 - 5.635 ] & K5V [ K5V - K3V ] \\ 
				Sh2-71~CS~1 & 0.86$\pm$0.07 & -0.342$\pm$0.292 & 0.17$\pm$0.292 & $>$8.655 & later than M3V \\ 

				&&&&&\\
				
				\hline
				PN Name	&	$E(B-V)$ & $(g-J)_{0}$	&	$\Delta(g-J)$ & M$_{J,2}$ & Comp. spec. type	\\
				\hline
				
				NGC~6337 & 0.48$\pm$0.03 & -0.420$\pm$0.141 & 0.599$\pm$0.141 & 6.486 [ 6.771 - 6.201 ] & M1V [ M2V - M1V] \\

				\hline
		\end{tabular}}
		\label{tab:excess_results}
	\end{center}
\end{table*}

\section{Individual Objects}
\label{sec:individual_pn}

\subsection{Hf~38}
\label{subsec:hf_38}

The new, superior resolution image from VPHAS+ has allowed us to identify the true CS of Hf~38. The object labelled throughout this work as Hf~38~CS~2 has a luminosity and colours consistent with a CS, whereas the colours of the off-centre star suggest that it is a main sequence star.

The $E(B-V)$ value calculated using the $u-g$ colour of CS 2 is larger than those derived from other methods. Given the close angular proximity of CS 1, CS 2 and the presence of a faint red star to the south of CS 2, it is likely that this is due to contamination. This object should be re-examined at a higher resolution to determine the true $E(B-V)$ value and search for $i$ band excess. 

While it is likely that CS 1 is a foreground star, it is also possible that CS 1 and CS 2 are a binary system.  Assuming the CS is at a distance of 2.25kpc \citep{frew2016}, the separation between the two stars is $\sim$5000 AU. On the main sequence, binary systems with this separation are observed, but are rare \citep{raghavan2010}.

\begin{figure}
	\centering
	\includegraphics[width=\linewidth]{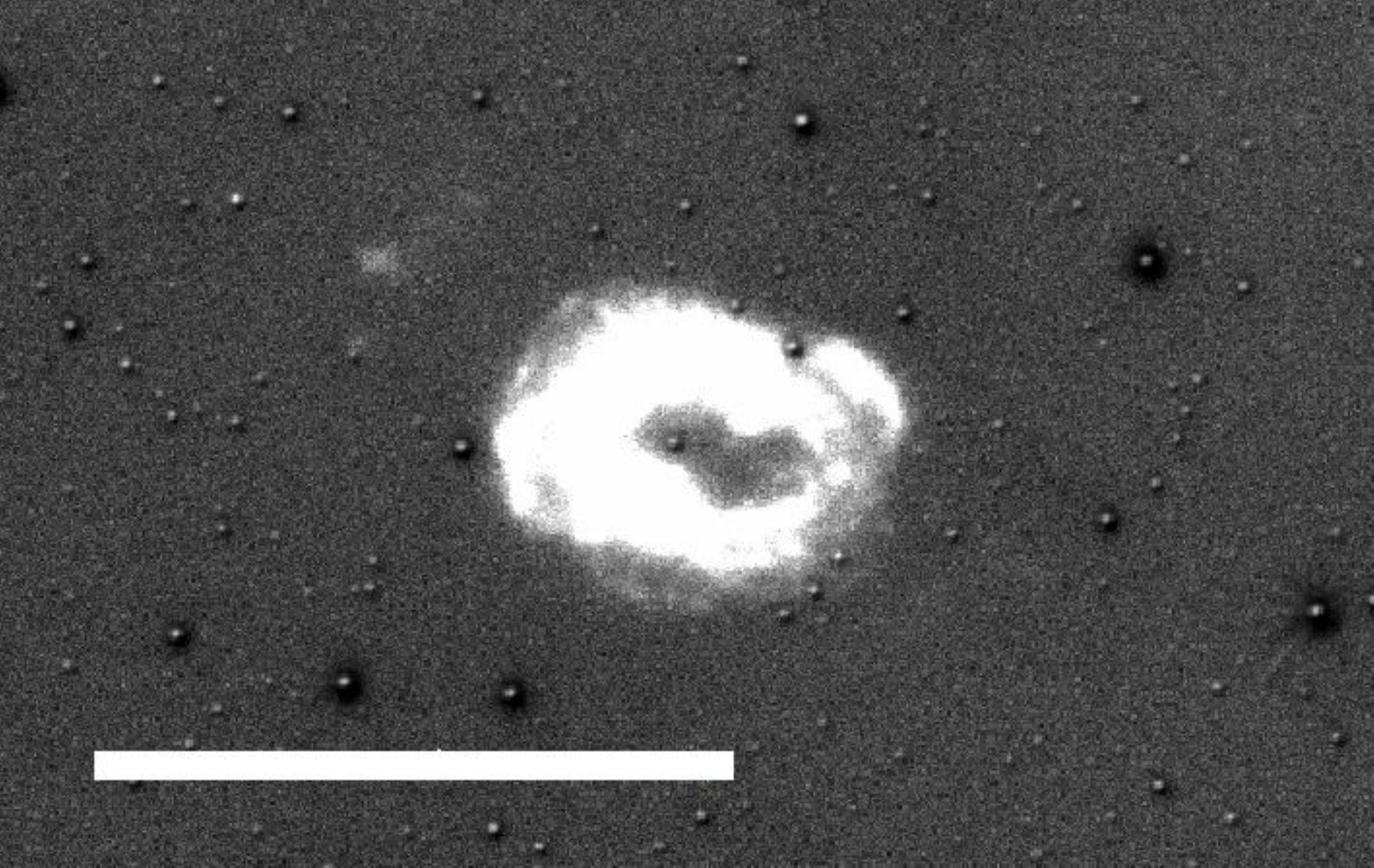}
	\caption{VPHAS+ image of Hf~38 in H$\alpha$/$r$ with north towards the top of the image, and a scale bar indicating one arcminute. Only CS 1 is visible. There is some nebulosity approximately 34 arcseconds north west of the CS.}
	\label{fig:Hf38_quotient_redo}
\end{figure}

We also present here the H$\alpha$/$r$ quotient image of Hf~38 from VPHAS+ in Figure \ref{fig:Hf38_quotient_redo}. There is some previously unseen nebulosity approximately 34 arcseconds north west of the CS, along the polar axis of the PN. Again, assuming the CS is at a distance of 2.25kpc, the distance from the CS to the nebulosity is around 74 000AU ($\sim$0.36 parsec).

\subsection{NGC~6337}
\label{subsec:NGC6337}

NGC 6337 is a known close binary with an orbital period of 0.17 days \citep{hillwig2010}, and is a pole-on elliptical PN \citep{corradi2000}. This large, bright nebula is well-studied, with several distance and temperature measurements in the literature, and so provides a good test of the $i$ band excess binary detection method used here.

\cite{hillwig2010} used time-resolved photometry and spectroscopy to model this binary system. They note the well-behaved brightness variability, also observable in the $r$ and $r2$ VPHAS+ measurements presented here, and therefore assume that the secondary star's radius is not greater than its Roche lobe, putting an upper mass limit on the secondary. They found the best agreement of their observations to their models using a CS temperature of $>$90kK and a main sequence companion with a mass of 0.35M$_{\odot}$, corresponding to a mid-M spectral type.

However, the irradiation of the secondary star was not fully accounted for in their models. Given the substantial variability of the light in the system, the mid-K spectral type from the $i$ band excess presented in this work is not counter to their findings.

\subsection{PNG~242.6-04.4}
\label{subsec:png242.6-04.4}

In the new, higher resolution VPHAS+ image we can see more clearly the `S'-shaped north-south outflow from the star at the centre of the nebula. This star is a good CS candidate based on its location, however on a colour-colour diagram, this appears to be a main sequence star.

Using the reddening map of \cite{green2015}, at the distance of this PN, 6.6$\pm$1.2 kpc, the sight-line has $E(B-V)$ = $0.47^{+0.033}_{-0.041}$. The $E(B-V)$ calculated for this PN using the CS $u-g$ colour is 1.77, suggesting the PN is very self-reddened. We suggest that the CS examined in this work is the true CS, but the light is dominated by that of a bright companion main sequence star, meaning the $E(B-V)$ calculated in this work is not the true value as our assumption that the CS light dominates the $u$ and $g$ bands does not hold. 

A crude blackbody analysis of this CS was performed in an attempt to constrain it's constituents. This analysis was inconclusive; no combination of model binary systems adequately reproduced the observed magnitudes and follow-up observation are required.

\subsection{PNG~288.2+00.4}
\label{subsec:png288.2+00.4}

Following inspection of the VPHAS+ and POPIPlaN images, we are confident that we have identified the true CS. However, the $g-i$ colour of this CS lies below the expectation value, the $u-g$ and $g-r$ colours are not definitively consistent with a true CS, and the estimated luminosity is very inconsistent with that of a true CS. We therefore suggest that there is considerable contamination, possibly due to a bright unresolved companion, causing the derived $E(B-V)$ value calculated using the CS colours to be too large. No other sources were available from which $E(B-V)$ could be calculated.

\subsection{PNG~293.4+00.1}
\label{PNG293.4+00.1}

This round, faint PN is ideal for the IR excess method presented here, and the blue star located in the centre of the PN is almost certainly the true CS. The CS has an $i$ band excess consistent with a mid-M spectral type companion star. Unfortunately, there are no $J$ band measurements in the literature with which to corroborate our detection.

\subsection{PTB~25}
\label{subsec:ptb25}

This PN was discovered by \cite{boumis2003} and simultaneously catalogued in the preliminary results of the MASH catalogue \citep{parker2006}. The field of this PN is very crowded, but a particularly blue star in the centre of the PN has been identified in VPHAS+ as the CS candidate. De-reddening using the $E(B-V)$ value derived from the $u-g$ colour, an $i$ band excess consistent with a mid-K spectral type star is detected. It remains unclear whether this is due to an unresolved binary companion, or contamination by neighbouring stars.

\subsection{Sh~2-71}
\label{subsec:SH2-71}

The true CS of Sh~2-71 has been the subject of debate. Historically, the 13.5 magnitude star labelled as CS~2 throughout this paper has been considered the CS (cf. \citealt{mocnik2015}). However, \cite{frew&parker2007} claim that the fainter star labelled here as CS~1, is the true CS. This object is closer to the geometric centre of the PN, and has magnitudes more consistent with that of a typical CS. 

The $E(B-V)$ calculations presented in Table \ref{tab:excess_results} support this, as the $E(B-V)$ value calculated using the $u-g$ colour of CS~1 is a far closer match to the those in literature (derived from observations of the nebula), than that of CS~2. This similarity in $E(B-V)$ values strongly suggests that both the nebula and CS~1 are at the same distance and are related. 

Nonetheless, a crude blackbody analysis was performed using the observed VPHAS+ magnitudes of CS~2 to determine if it could possibly be the true CS. The analysis performed assuming CS~2 comprises of a CS and a main sequence star, that it is at the distance of the nebula, 1.32 kpc \cite{frew2016}, and that it has a line of sight reddening similar to the nebula of $E(B-V)$=0.86, suggests that the star is a 140kK CS with an A0 companion. This is similar to the result of \cite{cuesta1993} who suggested the nebula hosts an ionising star with a temperature of 100kK and a A7V companion.

A second analysis was also performed $without$ the assumption that CS~2 contains a CS, and that it may be a foreground star superimposed on the nebula. Optimising over a grid of possible main sequence binaries, distances and reddenings, the best match to the observed magnitudes of CS~2 was a binary consisting of a F0V and a F5V star, at a distance of 800 pc with a line of sight reddening of $E(B-V)=0.6$. Using the method of determining $E(B-V)$ from the $u-g$ colour presented in this work, an observer would calculate the $E(B-V)$ of this system to be 1.54. This is less than the measured $E(B-V)=1.91$ calculated here, suggesting that if CS~2 is a foreground star, it is self-reddened.

Sh~2-71~CS~2 is known to vary in brightness by at least 0.7 magnitudes \citep{kohoutek1979} with a possible period of 68.132$\pm$ days \citep{mikulasek2005}. The VPHAS+ B and C block $g$ band measurements of CS~2 were taken approximately 730.047 days apart, giving a 0.715 period offset between measurements. The 0.527 $g$ band magnitude variation measured in this work is consistent with the known brightness variability. 

The $r$ and $r$2 VPHAS+ measurements were taken only 3 nights apart, but a difference of 0.529 magnitudes is detected in Sh~2-71~CS~1, an object with no previously observed variability. This difference in $r$ band measurements is far greater than other stars in the CCD, leading us to believe it is a real variability.

\section{Conclusion}
\label{sec:conclusion}

The goal of this work was to assess the potential of VPHAS+ to study PNe and their CS, and to determine if the photometry provided is sufficiently precise enough to detect CS IR excess indicative of a low mass main sequence companion. We conclude the following:

\begin{enumerate}
	\item The high-resolution images provided by VPHAS+ allows the identification of previously unobserved PN features, such as the internal nebula structure of PNG~242.6-04.4 presented here. The $u$ band images allows the identification of new CS candidates; a good example is that of PTB~25, which lies in a particularly crowded field.
	
	\item VPHAS+ has good seeing (with a median seeing of $\sim$0.8 arcseconds in $g$) and stable photometry, with the magnitudes of objects observed several times being consistent to within a few percent. These repeat measurements taken as part of the multi-offset observing plan also reduces the formal error on the magnitude measurements, and allows the identification of CS variability.
	
	\item Using the calibration method presented here, the typical systematic errors between APASS and VPHAS+ photometry using the 1.0 arcsecond aperture in the $g$, $r$, and $i$ bands is less than 0.01 magnitudes, with typical total formal errors on a single VPHAS+ $\sim$19$^{th}$ magnitude measurement less than 0.016, 0.025, and 0.044 magnitudes, respectively. The uncertainty is greater in the $i$ band as there is more noise associated with the photon count. These error decreases for objects with multiple measurements, and for the targets in this work, were generally considerably lower. The $u$ band calibration is generally less accurate, as there are no $u$ band magnitudes in APASS to calibrate to. Here, the typical systematic uncertainty using the 1.0 arcsecond aperture is 0.025 magnitudes, resulting in typical formal error on a single VPHAS+ $\sim$19$^{th}$ magnitude measurement of 0.048 magnitudes. It is hoped that as VPHAS+ completes (in 2019 at present rates of progress) and a final global calibration is put in place, APASS $g$, $r$ and $i$ data can be supplemented with that from other digital surveys such as PanSTARRS and Skymapper, supporting a more precise $u$ calibration. This will also reduce the number of objects lost due to insufficient calibration data. In this work, 7 of the initial 20 CS candidates had to be rejected from our sample due to poor APASS coverage.
	
	\item Despite the current challenges in calibrating the $u$ band, we find VPHAS+ photometry is sufficient for the highly precise IR excess analysis presented here. We detect an $i$ band excess in 3 of the 7 objects in our final sample. While follow-up observations are required to confirm CS binarity, excess flux was detected in the known close-binary NGC 6337 \citep{hillwig2006} in the $i$ band, and in the $J$ band using data from 2MASS, giving us confidence that our technique reliably detects CS companions.
	
	\item We also present evidence that the light of 2 CS in our sample is dominated by that of an unresolved main sequence companion, an indication of CS binarity.
	
\end{enumerate}	
	
The IR excess results of the trial sample of CS used in this work have not been used estimate the CS binary fraction. Rather, the work presented here highlights the value of the blue sensitivity of VPHAS+ for studying CS, and has shown that examination of the entire VPHAS+ catalogue to search for IR excess is viable. Given the large number of known PNe within the VPHAS+ footprint, this will likely provide a statistically significant number of IR excess-tested CS from which the rate of CS binarity can be determined. This work also highlights the need for near IR photometry of CS, as the $J$ band is more sensitive to low-mass binary companions but near IR sky-survey coverage is currently incomplete.

\section*{Acknowledgements} 

HB thanks Iain McDonald for his advice in data analysis, Christophe Morisset for assistance with pyCloudy, and acknowledges the STFC funding that made the visit to Macquarie University possible. ODM acknowledges funding from the Australian Research Council future fellowship (FT120100452). This research is based on data products from observations made with ESO Telescopes at the La Silla Paranal Observatory under programme ID 177.D-3023, as part of the VST Photometric Hα Survey of the Southern Galactic Plane and Bulge (VPHAS+, www.vphas.eu). JD acknowledges STFC consolidated grant support (ST/M001008/1). The TheoSSA service (http://dc.g-vo.org/theossa) used to retrieve theoretical spectra for this paper was constructed as part of the activities of the German Astrophysical Virtual Observatory. This research made use of Astropy, a community-developed core Python package for Astronomy (Astropy Collaboration, 2013). This research has made use of the HASH PN database, and the Centre de Données astronomiques de Strasbourg (CDS) database.

\bibliography{../Bibliography}

\clearpage
\appendix
\section{Synthetic colours in VPHAS+}
\label{app:individual magnitudes}

This appendix presents the expected colours of the central stars of PN and main sequence stars when observed using VPHAS+. The synthetic post-AGB atmospheres used are from TMAP calculated with the code TMAW \citep{rauch2003, werner1999, werner2003} or the German Astrophysical Virtual Observatory grid calculations TheoSSA\footnote{http://dc.zah.uni-heidelberg.de/theossa/q/web/form}. Also shown are the colours of a T=100kK, log(g)=7.0 synthetic atmosphere when reddened over a range of $E(B-V)$ values using the reddening laws of \cite{cardelli1989} and \cite{fitzptrick2007} with R$_v$=3.1. The spectra used to calculate main sequence star colours are from \cite{pickles1998}.

\begin{table*}
	\begin{center}
		\caption{Synthetic, intrinsic colours of CS using VPHAS+ bands, calculated using the theoretical stellar atmosphere models of TMAP. }
		\scalebox{0.8}
		{\begin{tabular}{l c c c c c c c}
				\hline
				T (kK) & log(g) & $u-g$ & $g-r$ & $r-i$ & $g-i$ & $g-J$ & Hydrogen/Helium fraction\\
				\hline
				20.0 & 4.0 & --1.040 & --0.191 & --0.099 & --0.290 & --0.583 & 0.741 \ 0.250 \\ 
				20.0 & 5.0 & --1.006 & --0.191 & --0.099 & --0.290 & --0.587 & 0.741 \ 0.250  \\
				30.0 & 4.0 & --1.347 & --0.267 & --0.141 & --0.408 & --0.813 & 0.741 \ 0.250  \\
				30.0 & 5.0 & --1.326 & --0.268 & --0.141 & --0.409 & --0.817 & 0.741 \ 0.250  \\
				30.0 & 7.0 & --1.337 & --0.261 & --0.137 & --0.398 & --0.820 & 0.712 \ 0.278  \\
				30.0 & 8.0 & --1.366 & --0.247 & --0.131 & --0.378 & --0.813 & 0.712 \ 0.278  \\
				40.0 & 7.0 & --1.494 & --0.295 & --0.157 & --0.452 & --0.926 & 0.712 \ 0.278  \\
				40.0 & 8.0 & --1.508 & --0.288 & --0.153 & --0.440 & --0.914 & 0.712 \ 0.278  \\
				50.0 & 6.0 & --1.534 & --0.297 & --0.157 & --0.454 & --0.922 & 0.712 \ 0.278  \\
				50.0 & 7.0 & --1.543 & --0.299 & --0.158 & --0.457 & --0.932 & 0.712 \ 0.278  \\
				60.0 & 6.0 & --1.558 & --0.306 & --0.161 & --0.467 & --0.945 & 0.712 \ 0.278  \\
				60.0 & 7.0 & --1.566 & --0.306 & --0.162 & --0.468 & --0.950 & 0.712 \ 0.278  \\
				70.0 & 6.0 & --1.573 & --0.311 & --0.163 & --0.474 & --0.957 & 0.712 \ 0.278  \\
				70.0 & 7.0 & --1.579 & --0.312 & --0.165 & --0.477 & --0.964 & 0.712 \ 0.278  \\
				80.0 & 7.0 & --1.593 & --0.318 & --0.168 & --0.486 & --0.979 & 0.712 \ 0.278  \\
				90.0 & 6.0 & --1.601 & --0.326 & --0.172 & --0.498 & --1.000 & 0.712 \ 0.278 \\
				90.0 & 7.0 & --1.605 & --0.325 & --0.171 & --0.496 & --0.997 & 0.712 \ 0.278 \\
				100.0 & 6.0 & --1.612 & --0.333 & --0.175 & --0.508 & --1.015 & 0.712 \ 0.278  \\
				100.0 & 7.0 & --1.616 & --0.332 & --0.174 & --0.506 & --1.011 & 0.712 \ 0.278  \\
				110.0 & 7.0 & --1.626 & --0.336 & --0.176 & --0.512 & --1.019 & 0.712 \ 0.278  \\
				120.0 & 7.0 & --1.633 & --0.339 & --0.177 & --0.516 & --1.025 & 0.712 \ 0.278  \\
				130.0 & 7.0 & --1.638 & --0.342 & --0.178 & --0.519 & --1.031 & 0.712 \ 0.278  \\
				140.0 & 7.0 & --1.642 & --0.343 & --0.178 & --0.522 & --1.034 & 0.712 \ 0.278 \\
				150.0 & 6.0 & --1.637 & --0.342 & --0.178 & --0.520 & --1.035 & 0.712 \ 0.278  \\
				150.0 & 7.0 & --1.646 & --0.345 & --0.179 & --0.524 & --1.037 & 0.712 \ 0.278 \\
				160.0 & 7.0 & --1.650 & --0.346 & --0.179 & --0.525 & --1.039 & 0.712 \ 0.278  \\
				170.0 & 7.0 & --1.654 & --0.348 & --0.180 & --0.528 & --1.045 & 0.712 \ 0.278 \\
				\hline
\end{tabular}}
\label{tab:synthetic_CS_colours}
\end{center}
\end{table*}

\begin{table*}
	\begin{center}
		\caption{Synthetic VPHAS+ colours of a CS, calculated using a synthetic stellar atmosphere from TMAP with a temperature of 100kK and log(g) = 0.7, reddened using the reddening laws of \protect\cite{cardelli1989} and \protect\cite{fitzptrick2007}.}
		\scalebox{0.8}
		{\begin{tabular}{l ccc ccc}
				\hline
				$E(B-V)$ & \multicolumn{3}{c}{\protect\cite{cardelli1989}} & \multicolumn{3}{c}{\protect\cite{fitzptrick2007}} \\
				 & $u-g$ & $g-r$ & $r-i$ & $ u-g$ & $g-r$ & $r-i$ \\
				\hline
				0.0	&	--1.616	&	--0.332	&	--0.174	&	--1.610	&	--0.330	&	--0.174  \\
				0.1	&	--1.504	&	--0.216	&	--0.113	&	--1.501	&	--0.202	&	--0.110\\
				0.2	&	--1.392	&	--0.103	&	--0.052	&	--1.392	&	--0.074	&	--0.045\\
				0.3	&	--1.281	&	0.009	&	0.009	&	--1.281	&	0.052	&	0.019\\
				0.4	&	--1.169	&	0.121	&	0.070	&	--1.170	&	0.178	&	0.082\\
				0.5	&	--1.056	&	0.231	&	0.131	&	--1.058	&	0.304	&	0.146\\
				0.6	&	--0.941	&	0.340	&	0.192	&	--0.945	&	0.428	&	0.210\\
				0.7	&	--0.826	&	0.448	&	0.253	&	--0.830	&	0.552	&	0.273\\
				0.8	&	--0.709	&	0.555	&	0.315	&	--0.715	&	0.675	&	0.336\\
				0.9	&	--0.591	&	0.662	&	0.376	&	--0.599	&	0.797	&	0.399\\
				1.0	&	--0.472	&	0.767	&	0.438	&	--0.482	&	0.918	&	0.461\\
				1.1	&	--0.352	&	0.871	&	0.499	&	--0.364	&	1.039	&	0.524\\
				1.2	&	--0.231	&	0.975	&	0.561	&	--0.246	&	1.159	&	0.586\\
				1.3	&	--0.109	&	1.077	&	0.623	&	--0.126	&	1.278	&	0.648\\
				1.4	&	0.015	&	1.179	&	0.685	&	--0.005	&	1.397	&	0.710\\
				1.5	&	0.139	&	1.280	&	0.746	&	0.116	&	1.514	&	0.771\\
				1.6	&	0.264	&	1.380	&	0.808	&	0.238	&	1.631	&	0.833\\
				1.7	&	0.390	&	1.479	&	0.871	&	0.360	&	1.748	&	0.894\\
				1.8	&	0.516	&	1.577	&	0.933	&	0.483	&	1.863	&	0.955\\
				1.9	&	0.642	&	1.675	&	0.995	&	0.605	&	1.978	&	1.016\\
				2.0	&	0.769	&	1.772	&	1.057	&	0.728	&	2.093	&	1.077\\
				2.1	&	0.896	&	1.868	&	1.120	&	0.850	&	2.206	&	1.138\\
				2.2	&	1.021	&	1.964	&	1.182	&	0.971	&	2.319	&	1.198\\
				2.3	&	1.146	&	2.059	&	1.245	&	1.090	&	2.432	&	1.258\\
				2.4	&	1.269	&	2.153	&	1.307	&	1.207	&	2.544	&	1.318\\
				2.5	&	1.389	&	2.247	&	1.370	&	1.319	&	2.655	&	1.378\\
				2.6	&	1.505	&	2.340	&	1.433	&	1.427	&	2.766	&	1.438\\
				2.7	&	1.615	&	2.432	&	1.495	&	1.527	&	2.876	&	1.498\\
				2.8	&	1.717	&	2.524	&	1.558	&	1.617	&	2.986	&	1.557\\
				2.9	&	1.81	&	2.616	&	1.621	&	1.696	&	3.095	&	1.617\\

				\hline
		\end{tabular}}
		\label{tab:synthetic_reddened_CS}
	\end{center}	
\end{table*}

\begin{table*}
	\begin{center}
		\caption{Synthetic intrinsic main sequence star colours using spectra from \protect\cite{pickles1998}.  M$_{V}$ values are from \protect\cite{demarco2013}. }
		\scalebox{0.8}
		{\begin{tabular}{l c c c c c c c c c}
				\hline
				Spectral Type & $u-g$ & $g-r$ & $r-i$ & $g-i$ & $i-g$ & $g-J$ & $M_{V}$ & $M_{J}$ & $M_{i}$ \\
				\hline
				
				O5V & 	--1.513 &--0.342 & --0.158 & 	--0.501 & 	0.501 & 	--0.966 & 	-	 & - & 	-	\\
				O9V	 & --1.439 & --0.343 & --0.135 & 	--0.479 & 	0.479 & 	--0.923 & 	- & 	- & 	-	\\
				B0V	 & --1.369 & --0.311	& --0.127 & 	--0.438 & 	0.438 & 	--0.897	 & - & 	-	& -	 \\
				B1V	 & --1.239 & --0.215 & --0.107 & 	--0.322 & 	0.322 & 	--0.779	 & - & 	- &	- 	\\
				B3V	 & ---0.936 & --0.185	& --0.043 & 	--0.229 & 	0.229 & 	--0.611 & 	--1.45 & 	--1.07 & 	--1.45	\\
				B8V	 & --0.372 & --0.061	& --0.020 & 	--0.081 & 	0.081 & 	--0.306 & 	--0.24 & 	--0.05 & 	--0.28	\\
				B9V	 & --0.266 & --0.011 & 	--0.016 & --0.027 & 	0.027 & 	--0.173 & 	0.16 & 	0.25 & 	0.10	\\
				A0V	 & --0.023 & 0.014 & 	0.026 & 	0.040 & 	--0.040 & 	0.004 & 	0.79 & 	0.74 & 	0.70	\\
				A2V	 & 0.046 & 	0.043 & 	0.041 & 	0.085 & 	--0.085 & 	--0.017 & 	1.36 & 	1.19 & 	1.09	\\
				A3V	 & 0.042 & 	0.115 & 	0.062 & 	0.177 & 	--0.177 & 	0.079 & 	1.53 & 	1.33 & 	1.23	\\
				A5V	 & 0.092 & 	0.166 & 	0.097 & 	0.263 & 	--0.263 & 	0.379 & 	1.90 & 	1.54 & 	1.66	\\
				A7V	 & 0.097 & 	0.244 & 	0.123 & 	0.367 & 	--0.367 & 	0.481 & 	2.16 & 	1.71 & 	1.82	\\	
				F0V	 & 0.057 & 	0.367 & 	0.188 & 	0.555 & 	--0.555 & 	0.714 & 	2.63 & 	2.03 & 	2.19	\\
				F2V	 & 0.017 & 	0.452 & 	0.227 & 	0.679 & 	--0.679 & 	0.837 & 	3.00 & 	2.28 & 	2.44	\\
				F5V	 & --0.011 & 0.519 & 	0.232 & 	0.751 & 	--0.751 & 	1.106 & 	3.46 & 	2.57 & 	2.93	\\	
				F6V	 & 0.031 & 	0.549 & 	0.271 & 	0.820 & 	--0.820 & 	1.187 & 	- & 	- & 	-	\\
				F8V	 & 0.114 & 	0.624 & 	0.287 & 	0.911 & 	--0.911 & 	1.339 & 	4.01 & 	2.95 & 	3.38	\\
				G0V	 & 0.189 & 	0.645 & 	0.326 & 	0.971 & 	--0.971 & 	1.355 & 	4.40 & 	3.28 & 	3.66	\\
				G2V	 & 0.264 & 	0.715 & 	0.340 & 	1.055 & 	--1.055 & 	1.590 & 	4.72 & 	3.54 & 	4.08	\\	
				G5V	 & 0.389 & 	0.748 & 	0.348 & 	1.097 & 	--1.097 & 	1.591 & 	5.07 & 	3.82 & 	4.31	\\
				G8V	 & 0.504 & 	0.835 & 	0.371 & 	1.206 & 	--1.206 & 	1.784 & 	5.51 & 	4.13 & 	4.71	\\
				K0V	 & 0.578 & 	0.885 & 	0.393 & 	1.277 & 	--1.277 & 	1.843 & 	5.89 & 	4.39 & 	4.96	\\
				K2V	 & 0.862 & 	1.037 & 	0.417 & 	1.454 & 	--1.454 & 	2.199 & 	6.37 & 	4.70 & 	5.45	\\
				K3V	 & 1.011 & 	1.124 & 	0.484 & 	1.608 & 	--1.608 & 	2.380 & 	6.61 & 	4.80 & 	5.57	\\
				K4V	 & 1.119 & 	1.268 & 	0.540 & 	1.808 & 	--1.808 & 	2.629 & 	- & 	- & 	-	\\
				K5V	 & 1.464 & 	1.440 & 	0.573 & 	2.013 & 	--2.013 & 	2.923 & 	7.34 & 	5.15 & 	6.06	\\
				K7V	 & 1.493 & 	1.599 & 	0.707 & 	2.306 & 	--2.306 & 	3.248 & 	8.16 & 	5.62 & 	6.56	\\
				M0V	 & 1.582 & 	1.589 & 	0.801 & 	2.390 & 	--2.390 & 	3.688 & 	8.87 & 	6.01 & 	7.31	\\
				M1V	 & 1.651 & 	1.590 & 	0.950 & 	2.540 & 	--2.540 & 	3.818 & 	9.56 & 	6.36 & 	7.64	\\
				M2V	 & 1.523 & 	1.698 & 	1.020 & 	2.718 & 	--2.718 & 	4.187 & 	10.17 & 	6.81 & 	8.28	\\
				M3V	 & 1.616 & 	1.741 & 	1.366 & 	3.108 & 	--3.108 & 	4.734 & 	11.01 & 	7.21 & 	8.84	\\
				M4V	 & 2.095 & 	1.795 & 	1.643 & 	3.438 & 	--3.438 & 	5.351 & 	12.80 & 	8.39 & 	10.30	\\
				M5V	 & 2.036 & 	1.898 & 	1.897 & 	3.795 & 	--3.795 & 	6.239 & 	14.20 & 	9.07 & 	11.51	\\
				M6V	 & 1.981 & 	2.061 & 	2.291 & 	4.351 & 	--4.351 & 	7.402 & 	16.59 & 	10.34 & 	13.39	\\

				\hline
		\end{tabular}}
		\label{tab:synthetic_MS_colours}
	\end{center}
\end{table*}

\section{Observational data from VPHAS+}
\label{app:vphas_obs}

Reported here are the VPHAS+ observation times and VPHAS+ single-band catalogue information, downloaded from CASU, of the objects discussed in this work.

\begin{table*}
	\begin{center}
		\caption{VPHAS+ single-band catalogue data, downloaded from CASU, and calibration information for our initial sample of CS candidates. The full table is available online. Observation times are in the format day/month/year hours:minutes:seconds, and magnitude correction entries are left empty for CCDs with a poor calibration.  }
		\scalebox{0.8}
		{\begin{tabular}{ll ll l l l l lllll l }
				\hline
				PN & Common & \multicolumn{2}{c}{Equatorial coordinates} & Aperture & VPHAS+ & Pointing  & CCD  & \multicolumn{5}{c}{Sequence numbers} & ... \\
				& name & RA & DEC & [arcsec] & pointing & block &number  & $u$ & $g$ & $r2$ & $r$ & $i$ & ...\\
				\hline
				
				\multirow{5}{*}{PNG 003.4+01.4} & \multirow{5}{*}{-} & \multirow{5}{*}{17:48:15.558} & \multirow{5}{*}{--25:15:14.34} & \multirow{5}{*}{1.41} & 0699 & A & 24 & 732 & 1346 & 4938 & 5879 & 11458 & ... \\
				& & & & & & & & & & & & & \\
				& & & & & \multirow{3}{*}{0700} & A & 5 & 260 & 755 & 3652 & 4020 & 7993 & ... \\
				& & & & & & B & 2 & 364 & 1206 & 6199 & 7003 & 13775 & ... \\
				& & & & & & C & 5 & - & 483 & - & - & - &... \\
				
				\multicolumn{14}{l}{}\\
				\multirow{3}{*}{PNG 018.8-01.9}  &\multirow{ 3}{*}{PTB 25} & \multirow{ 3}{*}{18:32:04.549} & \multirow{ 3}{*}{--13:26:15.08} & \multirow{3}{*}{1.0} & \multirow{ 3}{*}{0408} & A & 25 & - & 7602 & 11604 & 12790 & 13126 & ...\\
				& & & & & & B & 23 & 2870 & 12344 & 19341 & 18480 & 18268 & ...\\
				& & & & & & C & 26 & - & 755 & - & - & - &... \\
				
				\multicolumn{14}{l}{}\\
				\multirow{5}{*}{PNG 029.0+00.4} & \multirow{5}{*}{Abell 48} & \multirow{5}{*}{18:42:46.920} & \multirow{5}{*}{-03:13:17.24} &\multirow{5}{*}{2.0}  & \multirow{3}{*}{0136} & A & 2 & 388 & 2883 & 6769 & 5075 & 8237 & ... \\
				& & & & & & B & 3 & 4 & 35 & 125 & 179 & 306 & ... \\
				& & & & & & C & 2 & - & 1839 & - & - & - & ... \\
				& & & & & & & & & & & & & \\
				& & & & & 0162 & B & 15 & 513 & 2898 & 6977 & 6265 & 9144 & ... \\
				
				\multicolumn{14}{l}{}\\
				\multirow{ 2}{*}{PNG 035.9-01.1 CS 1} & \multirow{ 2}{*}{Sh~2-71 CS 1} & \multirow{ 2}{*}{19:01:59.954} & \multirow{ 2}{*}{02:09:16.17} & \multirow{2}{*}{1.41} & \multirow{ 2}{*}{0113} & B & 11 & 47 & 105 & 262 & 193 & 333 & ... \\
				& & & & & & C & 10 & - & 374 & - & - & - & ... \\
				
				\multicolumn{14}{l}{}\\
				\multirow{ 2}{*}{PNG 035.9-01.1 CS 2} & \multirow{ 2}{*}{Sh~2-71 CS 2} & \multirow{ 2}{*}{19:02:00.289} & \multirow{ 2}{*}{02:09:10.96} & \multirow{2}{*}{2.0}  & \multirow{ 2}{*}{0113} & B & 11 & 44 & 106 & 491 & 331 & 334 & ... \\
				& & & & & & C & 10 & - & 372 & - & - & - & ... \\
				
				\multicolumn{14}{l}{}\\
				\multirow{2}{*}{PNG 242.6-04.4} & \multirow{2}{*}{-} & \multirow{ 2}{*}{07:32:17.547} & \multirow{ 2}{*}{-28:25:17.92} & \multirow{2}{*}{1.41} & \multirow{ 2}{*}{0757} & A & 2 & 889 & 1409 & 2222 & 2392 & 2487 & ... \\
				& & & & & & C & 3 & - & 670 & - & - & - & ...\\
				
				\multicolumn{14}{l}{}\\
				\multirow{2}{*}{PNG~285.4-01.1} & \multirow{2}{*}{Pe~2--5} & \multirow{ 2}{*}{10:28:34.610} & \multirow{ 2}{*}{--59:03:23.28} & \multirow{ 2}{*}{2.0} & \multirow{ 2}{*}{1734} & A & 22 & 2833 & 7038 & 11196 & 8793 & 9350 & ... \\
				& & & & & & B & 24 & 379 & 905 & 2332 & 1127 & 1365 & ...\\
				& & & & & & C & 23 & - & 4132 & - & - & - & ... \\
				
				\multicolumn{14}{l}{}\\
				\multirow{ 2}{*}{PNG 288.2+00.4} & \multirow{ 2}{*}{-} & \multirow{ 2}{*}{10:53:31.977} & \multirow{ 2}{*}{--59:03:02.77} & \multirow{2}{*}{1.0} & \multirow{ 2}{*}{1738} & A & 5 & 2773 & 7434 & 11982 & 6685 & 12546 & ...\\
				& & & & & & B & 7 & 499 & 1289 & 2129 & 1411 & 2094 & ... \\
				& & & & & & C & 6 & - & 4732 & - & - & - & ... \\
				
				\multicolumn{14}{l}{}\\
				\multirow{2}{*}{PNG 288.4+00.3 CS 1} & \multirow{2}{*}{Hf~38 CS 1} & \multirow{2}{*}{10:54:35.456} & \multirow{2}{*}{--59:07:46.92} & \multirow{2}{*}{1.0} & \multirow{2}{*}{1738} & B & 4 & 1612 & 5333 & 9311 & 5896 & 9368 & ...\\
				& & & & & & C & 7 & - & 612 & - & - & - & ... \\
				
				\multicolumn{14}{l}{}\\
				\multirow{2}{*}{PNG 288.4+00.3 CS 2} &\multirow{2}{*}{Hf~38 CS 2} & \multirow{2}{*}{10:54:35.170} & \multirow{2}{*}{--59:09:46.67} &  \multirow{2}{*}{1.0} & \multirow{2}{*}{1738} & B & 4 & 1613 & 5335 & 9303 & 5894 & 9370 & ... \\
				& & & & & & C & 7 & -  & 611 & - & - & - & ...\\

				\multicolumn{14}{l}{}\\
				PNG 293.4+00.1 & - & 11:30:58.948 & --61:15:50.96 & 1.0 & 1884 & B & 30 & 2884 & 8359 & 13837 & 12573 & 14573 & ... \\

				
				\multicolumn{14}{l}{}\\
				\multirow{ 2}{*}{PNG 344.4+01.8} & \multirow{ 2}{*}{-} & \multirow{ 2}{*}{16:54:43.326} & \multirow{ 2}{*}{--40:41:47.05} & \multirow{2}{*}{2.0} & \multirow{ 2}{*}{1111} & A & 6 & 938 & 3065 &  7050 & 6635 & 10394 & ... \\
				& & & & & & B & 8 & 39 & 103 & 278 & 254 & 459 & ... \\
				& & & & & & C & 7 & - & 1812 & - & - & - & ... \\

				\multicolumn{14}{l}{}\\
				\multirow{ 3}{*}{PNG 349.3-01.1} & \multirow{ 3}{*}{NGC 6337} & \multirow{ 3}{*}{17:22:15.673} & \multirow{ 3}{*}{--38:29:01.67} &  \multirow{3}{*}{2.0} & \multirow{3}{*}{1061} & A & 12 & 1139 & 3755 &  6988 & 7400 & 11794  ...\\   
				& & & & & & B & 25 & 202 & 555 & 1102 & 1098 & 1498 & ... \\
				& & & & & & C & 12 & - & 1787 & - & - & - & ... \\
				
				\multicolumn{14}{l}{}\\
				\multirow{3}{*}{PNG 354.8+01.6} & \multirow{3}{*}{-} & \multirow{3}{*}{17:26:25.801} & \multirow{3}{*}{-32:21:49.84} & \multirow{3}{*}{2.0} & 0874 & A & 17 & 110 & 209 & 826 & 846 & 2743 & ...\\
				& & & & & & & & & & & & &\\
				& & & & & 0900 & B & 31 & 89 & 224 & 836 & 456 & 1769 & ... \\
				
				\multicolumn{14}{l}{}\\
				\multirow{4}{*}{PNG 355.9+00.7} & \multirow{4}{*}{-} &\multirow{4}{*}{17:32:44.493} & \multirow{4}{*}{--32:01:07.44}  & \multirow{4}{*}{1.41} & \multirow{2}{*}{0876} & A & 5 & 37  & 2012 &  4196 & 3286 & 5535 & ... \\
				& & & & & & C & 5 & - & 1004 & - & - & - & ... \\
				& & & & & & & & & & & & & \\
				& & & & & 0875 & A & 24 & 426 & 1751 & 4363 & 4522 & 6969 & ...\\
				
				\hline
		\end{tabular}}
		\label{tab:input_cat_data}
	\end{center}
\end{table*}

\section{Rejected CS candidates}
\label{ap:rejected_photometry}

Reported in this appendix is the VPHAS+ photometry of objects rejected from our CS sample, as we do not believe that they are the true CS. The C block $g$ band exposure of Abell~48, and A block measurements of PNG~344.4+01.8  in all bands were not used, as the calibration to APASS was poor.

\begin{table*}
	\begin{center}
		\caption{The calculated single-band VPHAS+ magnitudes for objects rejected from our CS sample. The number of measurements is shown in brackets, and the average magnitude quoted where multiple measurements were available. $J$ band measurements are from the literature.}
		\scalebox{0.8}
		{\begin{tabular}{lc ccccc c}
				\hline
				PN & Common Name & $u$ & $g$ & $r2$ & $r$ & $i$ & $J$ \\
				\hline		
				
				PNG 003.4+01.4 & - & 20.822$\pm$0.065 (3) & 18.770$\pm$0.009 (4) & 16.925$\pm$ 0.004 (3) & 16.928$\pm$0.004 (3) & 15.767$\pm$0.008 (3) & 13.791$\pm$0.005$^2$ \\				
				
				PNG 285.4-01.1 & Pe 2--5 & 15.798$\pm$0.029 (2) & 15.608$\pm$0.029 (3) & 14.889$\pm$0.011 (2) & 14.866$\pm$0.020 (2) & 14.887$\pm$0.015 (2) & 13.969$\pm$0.047 \\				
				
				PNG 288.4+00.3 CS 1 & Hf~38 CS 1 & 19.124$\pm$0.028 (1) & 18.016$\pm$0.009 (2) & 16.799$\pm$0.009 (1) &  16.729$\pm$0.014 (1) & 16.246$\pm$0.014 (1) & 15.069$\pm$0.087$^1$ \\

				PNG 344.4+01.8 & - & 20.388$\pm$0.077 (1) & 18.903$\pm$0.020 (2) & 17.255$\pm$0.023 (1) & 17.282$\pm$ 0.018 (1) & 16.557$\pm$0.037 (1) & 15.101$\pm$0.075$^1$ \\

				PNG 354.8+01.6 & - & 17.490$\pm$0.035 (2) & 16.520$\pm$0.003 (2) & 15.124$\pm$0.004 (2) & 15.129$\pm$0.004 (2) & 14.114$\pm$0.005 (2) & 12.761$\pm$0.002$^2$\\
				
				PNG 355.9+00.7 & - & 19.662$\pm$0.046 (2) & 20.488$\pm$0.035 (3) & 18.673$\pm$0.014 (2) & 18.821$\pm$0.017 (2) & 17.264$\pm$0.012 (2) & 14.068$\pm$0.068$^1$ \\
				
				PNG 029.0+00.4 & Abell 48 & 20.160$\pm$0.083 (3) & 19.244$\pm$0.015 (3) & 17.162$\pm$0.009 (3) & 17.127$\pm$0.007 (3) & 15.679$\pm$0.007 (3) & - \\

				\hline
				
				\multicolumn{7}{l}{$^1$ 2MASS } \\
				\multicolumn{7}{l}{$^2$ VVV }	\\

		\end{tabular}}
		\label{tab:rejected_input_mags}
	\end{center}
\end{table*}

\end{document}